\documentclass[aps,pra,twocolumn,floatfix,superscriptaddress,showpacs]{revtex4}
\usepackage{graphicx}
\usepackage{dcolumn}
\usepackage{bm}
\usepackage{amssymb,bezier,epsfig,amsmath}
\input rotate
\input colordvi
\begin{document}
\newcommand{\App}{A^{\prime\prime}}

\title{Cold collisions of OH and Rb. I: the free collision}
\author{Manuel Lara} \email{laram@murphy.colorado.edu}
\affiliation{JILA, University of Colorado, Boulder, CO 80309-0440}
\author{John L. Bohn} \email{bohn@murphy.colorado.edu}
\affiliation{JILA and Department of Physics, University of
Colorado, Boulder, CO 80309-0440}
\author{Daniel E. Potter}
\affiliation{Department of Chemistry, University of Durham,
England}
\author{Pavel Sold\'an} \email{pavel.soldan@fjfi.cvut.cz}
\affiliation{Department of Chemistry, Faculty of Nuclear Science
and Physical Engineering, Czech Technical University, Czech
Republic}
\author{Jeremy M. Hutson} \email{J.M.Hutson@durham.ac.uk}
\affiliation{Department of Chemistry, University of Durham,
England}

\date{\today}

\begin{abstract}
We have calculated elastic and state-resolved inelastic cross
sections for cold and ultracold collisions in the Rb($^1 S$) +
OH($^2 \Pi_{3/2}$) system, including fine-structure and hyperfine
effects. We have developed a new set of five potential energy
surfaces for Rb-OH($^2 \Pi$) from high-level {\em ab initio}
electronic structure calculations, which exhibit conical
intersections between covalent and ion-pair states. The surfaces
are transformed to a quasidiabatic representation. The collision
problem is expanded in a set of channels suitable for handling the
system in the presence of electric and/or magnetic fields,
although we consider the zero-field limit in this work. Because of
the large number of scattering channels involved, we propose and
make use of suitable approximations. To account for the hyperfine
structure of both collision partners in the short-range region we
develop a frame-transformation procedure which includes most of
the hyperfine Hamiltonian. Scattering cross sections on the order
of $10^{-13}$ cm$^2$ are predicted for temperatures typical of
Stark decelerators. We also conclude that spin orientation of the
partners is completely disrupted during the collision.
Implications for both sympathetic cooling of OH molecules in an
environment of ultracold Rb atoms and experimental observability
of the collisions are discussed.

\end{abstract}


\maketitle
\section{Introduction}

The possibility of producing translationally ultracold molecules
has recently generated great anticipation in the field of
molecular dynamics. Attractive applications include the
possibility of testing fundamental symmetries
\cite{EDM}, the potential of new phases of matter \cite{10, 11a,
11b, 12} and the renewed quest for the control of chemical
reactions through ultracold chemistry \cite{Gian, Balak, Bala}.
These endeavors are beginning to bear scientific fruit. For
example, high-resolution spectroscopic measurements of
translationally cold samples of OH should allow improved
astrophysical tests of the variation with time of the
fine-structure constant \cite{maser}. The recent experimental
advances \cite{Cold, Special, Form} have made theoretical studies
of the collisional behavior of cold molecules essential
\cite{Krems, Weck, Bodo}, both to interpret the data and to
suggest future directions.

Several approaches have produced cold neutral molecules to date,
many of which are described in Ref.\ \onlinecite{Special}.
The methods available can be classified into {\em direct methods},
based on cooling of preexisting molecules, and {\it indirect
methods}, which create ultracold molecules from ultracold atoms.
Among the direct methods, Stark deceleration of dipolar molecules
in a supersonic beam
\cite{4a, 4b, Boch} and helium buffer-gas cooling
\cite{3} are currently leading the way.
They reach temperatures of the order of 10~mK, and there are a
wide variety of proposals on how to bridge the temperature gap to
below 1~mK. These include evaporative cooling and even direct
laser cooling. The idea of {\em sympathetic cooling}, where a hot
species is cooled via collisions with a cold one, also seems very
attractive and is being pursued by several experimental groups. 
Sympathetic cooling is a form of {\em collisional cooling}, which
works for multiple degrees of freedom simultaneously. It does not
rely on specific transitions, which makes if suitable for cooling
molecules. Collisional cooling is also the basis for helium
buffer-gas cooling.

Sympathetic cooling of trapped ions has already been demonstrated
\cite{iones}, using a different laser-cooled ionic species as the
refrigerant. Cooling of polyatomic molecules to sub-kelvin
temperatures with ions has also been reported. This technique is
expected to be suitable for cooling molecules of very high mass,
including those of biological relevance \cite{biolo}. But the ease
with which alkali metal atoms can be cooled to ultracold
temperatures makes them good candidates to use as a thermal
reservoir to cool other species. They have already been used to
cool ``more difficult'' atomic alkali metal partners. For example,
BEC for $^{41}$K was achieved by sympathetic cooling of potassium
atoms with Rb atoms \cite{Modugno}. A
theoretical study of the viability of this cooling technique for molecules
is desirable.
There have been a number of theoretical studies of collisions of
molecules with He \cite{CaH, NH, OH, Forr, otra, what}, in support of buffer-gas
cooling, but only a few with alkali metals \cite{Na2, Li2homo,
Li2het, K2, PRL}. To our knowledge, no such study has included the
effects of hyperfine structure.

The main objective of the present work is to study cold collisions
of OH with trapped Rb atoms. OH has been successfully slowed by
Stark deceleration in at least two laboratories \cite{Meer,maser}.
 To cool the molecules further, sympathetic cooling by
thermal contact with $^{87}$Rb is an attractive possibility. Rb is
easily cooled and trapped in copious quantities and can be
considered the workhorse for experiments on cold atoms.
Temperatures below 100~$\mu$K are reached even in a MOT
environment (70~$\mu$K using normal laser cooling and 7~$\mu$K
using techniques such as polarization gradient cooling).

The cooling and lifetime of species in the trap depends largely on
the ratio of elastic collision rates (which lead to thermalization
of the sample) to inelastic ones. The latter can transfer
molecules into non-trappable states and/or release kinetic energy,
with resultant heating and trap loss. The characterization of the
rates of both kinds of process is thus required. Since applied
electric and magnetic fields offer the possibility of {\em
controlling} collisions, it is very important to know the effects
of such fields on the rates.
At present, nothing is known about the low-temperature collision
cross sections of Rb-OH or any similar system.

Rb-OH can be considered as a benchmark system for the study of the
 feasibility
of sympathetic cooling for molecules.
Many molecule-alkali metal atom systems have deeply bound
electronic states with ion-pair character \cite{17,9} and have
collision rates that are enhanced by a ``harpooning'' mechanism.
 Both the atom and the diatom are open-shell
doublet species, and can interact on two triplet and three singlet
potential energy surfaces (PES). In addition, the OH radical has fine
structure, including lambda-doubling, and both species possess
nuclear spins and hence hyperfine structure. Thus Rb-OH is
considerably more complicated than other collision systems that
have been studied at low temperatures. In previous work we
advanced the first estimates of cross sections (for both inelastic
and elastic collisions), based on fully {\em ab initio} surfaces,
for the collision of OH radicals with Rb atoms in the absence of
external fields \cite{PRL}. Here we provide details of the
methodology used and discuss the potential surfaces and the
state-resolved partial cross sections.

This paper is organized as follows: Section II describes the
calculation of {\em ab initio} PES for
Rb-OH. Details of the electronic structure calculations are given
and the methods used for diabatization, interpolation and fitting
are described. The general features of the resulting surfaces are
analyzed. Section III describes the exact and approximate
theoretical methodologies used for the dynamical calculations.
Section IV presents the resulting cross sections and discusses the
possibility of sympathetic cooling. We also comment on the role
expected for the harpooning mechanism. Section 5 summarizes our
results and describes prospects for future work.
Further details about electronic structure calculations and other 
 channels basis sets to describe the dynamics
are given in Appendixes 1 and 2 
respectively.

\section{Potential energy surfaces}

We have used {\it ab initio} electronic structure calculations to
obtain PES for interaction of OH($^2\Pi$)
with Rb. The ground $X^2\Pi$ state of OH has a $\pi^3$
configuration, while the ground $^2S$ state of Rb has 5s$^1$. At
long range, linear RbOH thus has $^1\Pi$ and $^3\Pi$ states. At
nonlinear configurations, $^1\Pi$ splits into $^1A'$ and $^1\App$,
with even and odd reflection symmetry in the molecular plane,
whereas $^3\Pi$ splits into $^3A'$ and $^3\App$.

\begin{figure}
\setlength{\unitlength}{4mm}
\begin{picture}(0,0)(0,0)
\put(8,-12){\makebox(0,0){{\bf \large{(a)}}}}
\put(8,-28.7){\makebox(0,0){{\bf \large{(b)}}}}
\put(8,-45.5){\makebox(0,0){{\bf \large{(c)}}}}
\end{picture}
\vspace{-.5cm}
\begin{center}

\epsfig{file=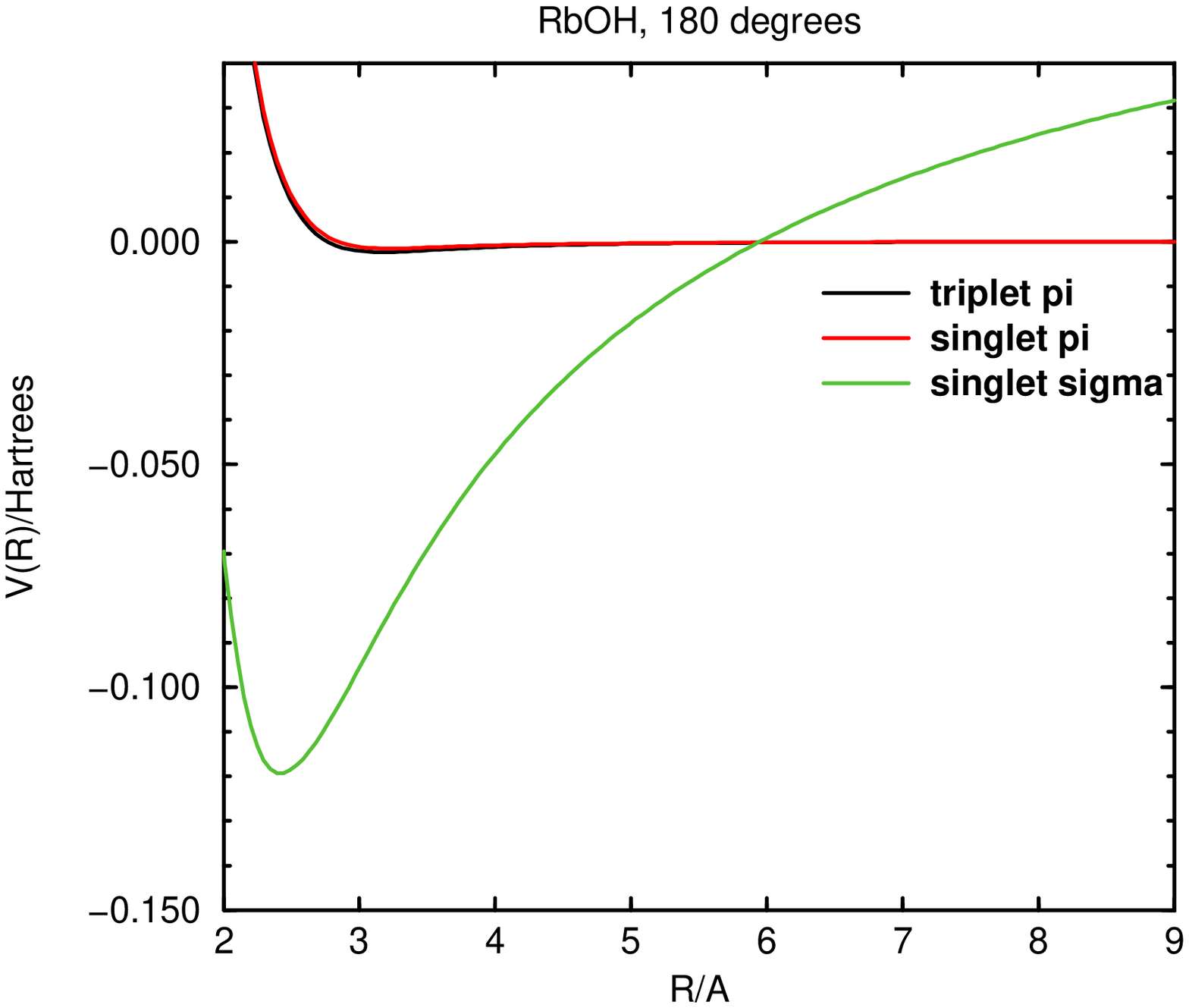,angle=0,width=0.95\linewidth,clip=}
\epsfig{file=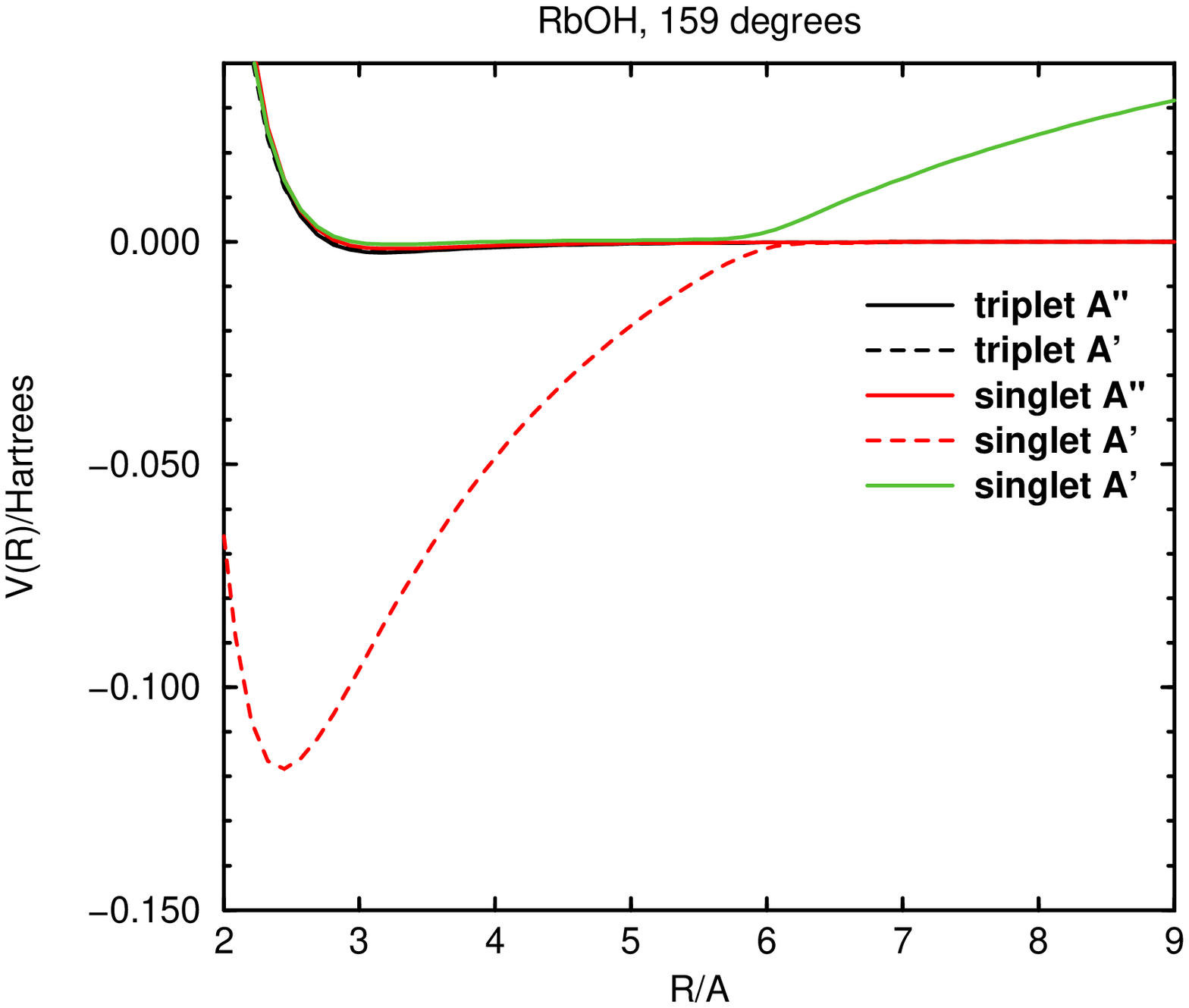,angle=0,width=0.95\linewidth,clip=}
\epsfig{file=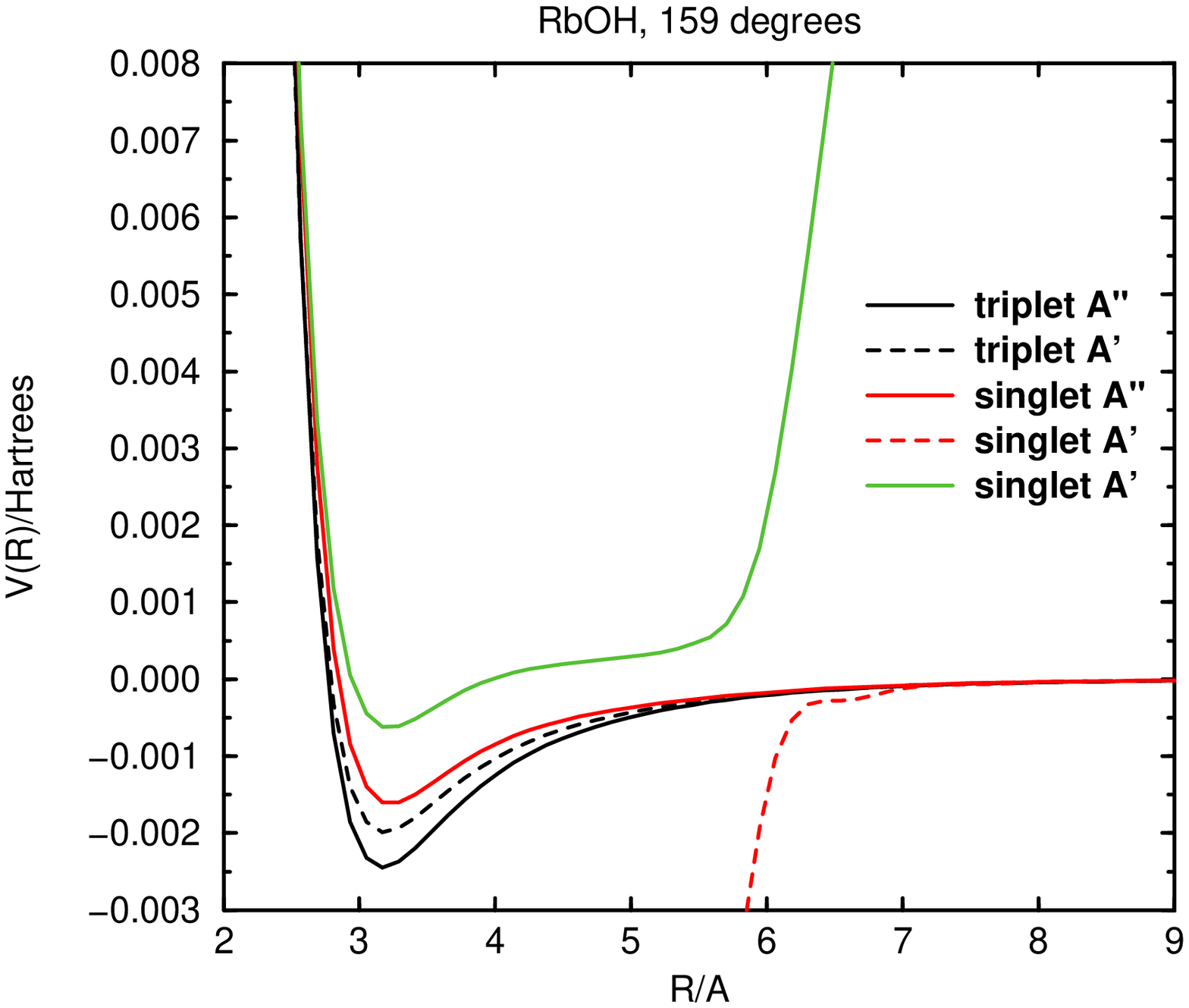,angle=0,width=0.95\linewidth,clip=}
\end{center}

\caption{RbOH adiabatic potential curves from MRCI calculations,
showing crossing for the Rb-OH linear geometry (a) and avoided
crossing for a slightly nonlinear geometry ($\theta=159^\circ$,
(b)). Panel (c) shows an expanded view of the curves in the center
panel.} \label{figcurve}
\end{figure}

At shorter range, the situation is more complicated. The ion-pair
threshold Rb$^+$ + OH$^-$ lies only 2.35 eV above the neutral
threshold. The corresponding $^1\Sigma^+$ ($^1A'$) ion-pair state
drops very fast in energy with decreasing distance because of the
Coulomb attraction. Below, Jacobi coordinates ($R,\theta$) will be used,
 $R$ being  the radial distance
between the atom and the OH center of mass and $\theta$ the
angle between this line and the internuclear axis. 
At the linear Rb-OH geometry, the ion-pair
state crosses the covalent (non-ion-pair) state near $R=6.0$ \AA,
as shown in Figure \ref{figcurve}. At nonlinear geometries, the
ion-pair state has the same symmetry ($^1A'$) as one of the
covalent states, so there is an avoided crossing. There is thus a
conical intersection between the two $^1A'$ states at linear
geometries, which may have major consequences for the scattering
dynamics.

The $^1A'$ electronic wavefunctions near the conical intersection
are made up from two quite different configurations, so that a
multiconfiguration electronic structure approach is essential to
describe them. We have therefore chosen to use MCSCF
(multiconfiguration self-consistent field) calculations followed
by MRCI (multireference configuration interaction) calculations to
characterize the surfaces. The electronic structure calculations
initially produce {\em adiabatic} (Born-Oppenheimer) surfaces, but
these are unsuitable for dynamics calculations both because they
are difficult to interpolate (with derivative discontinuities at
the conical intersections) and because there are nonadiabatic
couplings between them that become infinite at the conical
intersections. We have therefore transformed the two $^1A'$
adiabatic surfaces that cross into a {\it diabatic}
representation, where there are non-zero couplings between
different surfaces but both the potentials and the couplings are
smooth functions of coordinates.

The electronic structure calculations are carried out using the
MOLPRO package \cite{MOLPRO}. It was necessary to carry out an RHF
(restricted Hartree-Fock) calculation to provide initial orbital
guesses before an MCSCF calculation. It is important that the
Hartree-Fock calculation gives good orbitals for both the OH $\pi$
and Rb 5s orbitals at all geometries (even those inside the
crossing, where the Rb 5s orbital is unoccupied in the ground
state). In addition, it is important that the OH $\pi$ orbitals
are doubly occupied in the RHF calculations, as otherwise they are
non-degenerate at linear geometries at the RHF level, and the
MCSCF calculation is unable to recover the degeneracy. To ensure
this, we begin with an RHF calculation on RbOH$^-$ rather than
neutral RbOH.

For Rb, we use the small-core quasirelativistic effective core
potential (ECP) ECP28MWB \cite{ECPs} with the valence basis set from
Ref.\ \cite{Sol03}. This treats the 4s, 4p and 5s electrons
explicitly, but uses a pseudopotential to represent the core
orbitals. For O and H, we use the aug-cc-pVTZ correlation-consistent
basis sets of Dunning \cite{Dunning} in uncontracted form.
Electronic structure calculations were carried out at 275
geometries, constructed from all combinations of 25 intermolecular
distances $R$ and 11 angles $\theta$ in Jacobi coordinates. The 25
distances were from 2.0 to 6.0\,\AA\ in steps of 0.25\,\AA, from 6.0
to 9.0\,\AA\ in steps of 0.5\,\AA\ and from 9\,\AA\ to 12\,\AA\ in
steps of 1\,\AA. The OH bond length was fixed at $r=0.9706$\,\AA.
The 11 angles were chosen to be Gauss-Lobatto quadrature points
 \cite{Lobatto}, which give optimum quadratures to project out the
Legendre components of the potential while retaining points at the
two linear geometries. The linear points are essential to ensure
that the $A'$ and $\App$ surfaces are properly degenerate at
linear geometries: if we used Gauss-Legendre points instead, the
values of the $A'$ and $\App$ potentials at linear geometries
would depend on extrapolation from nonlinear points and would be
non-degenerate. The Gauss-Lobatto points correspond approximately
to $\theta=0$, 20.9, 38.3, 55.6, 72.8, 90, 107.2, 124.4, 141.7,
159.1 and 180$^\circ$, where $\theta=0$ is to the linear Rb-HO
geometry. The calculations were in general carried out as angular
scans at each distance, since this avoided most convergence
problems due to sudden changes in orbitals between geometries.

\subsection{Singlet states}

\begin{figure}
\begin{center}
\epsfig{file=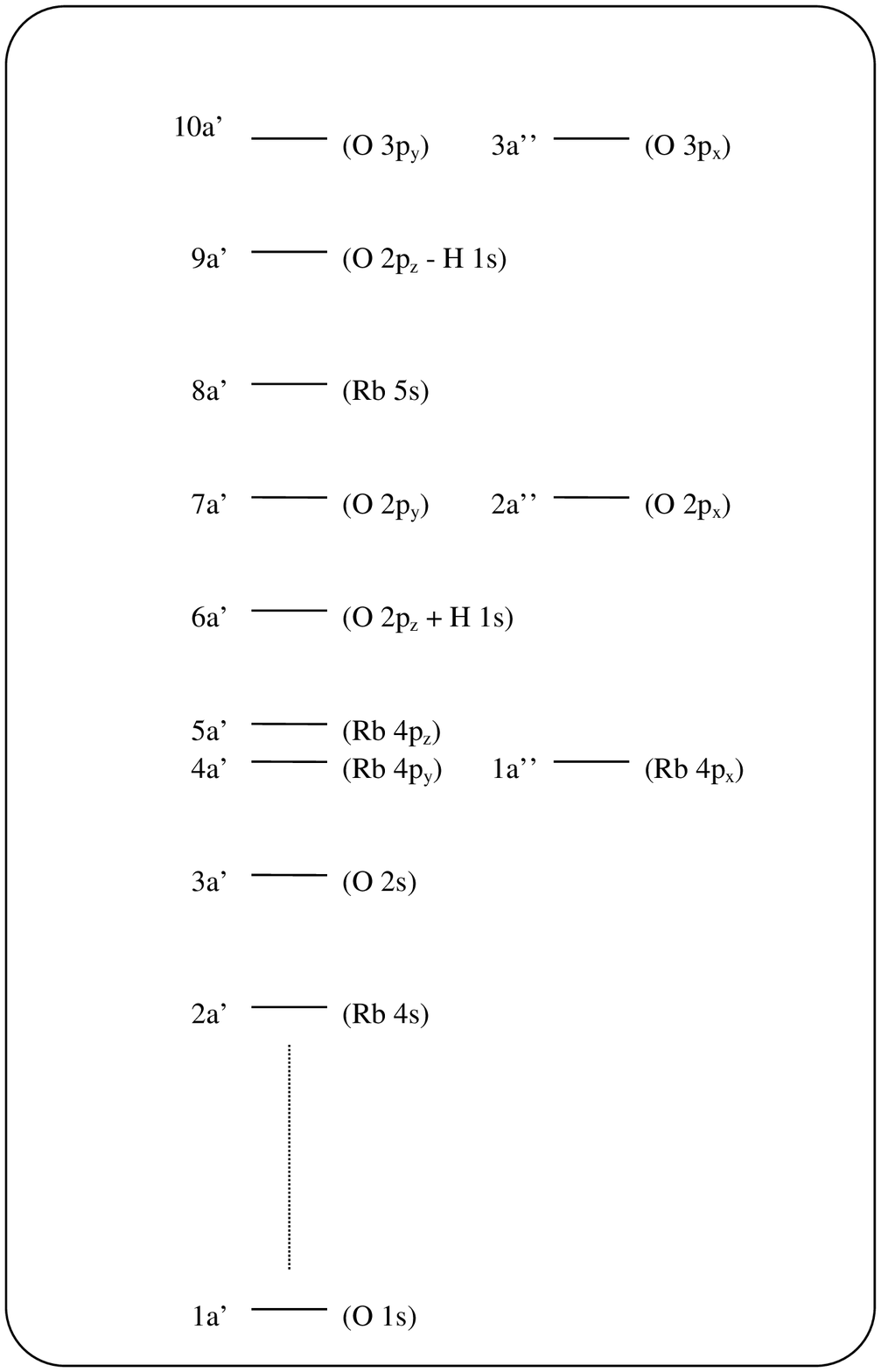,angle=0,width=85mm}
\end{center}
\caption{Schematic representation of the RbOH molecular orbitals
from MCSCF calculations.} \label{figorb}
\end{figure}

We carried out a state-averaged MCSCF calculation of the lowest 3
singlet states of neutral RbOH (two $^1A'$ and one $^1\App$).
Molecular orbital basis sets will be described using the notation
$(n_{A'},n_{\App})$, where the two integers indicate the number of
$A'$ and $\App$ orbitals included. The orbital energies are shown
schematically in Figure \ref{figorb}. The MCSCF basis set includes
a complete active space (CAS) constructed from the lowest (10,3)
molecular orbitals, with the lowest (5,1) orbitals closed (doubly
occupied in all configurations). The MCSCF calculation generates a
common set of orbitals for the 3 states. The calculations were
carried out in $C_s$ symmetry, but at linear geometries the two
components of the $\Pi$ states are degenerate to within our
precision ($10^{-8}\,E_h$).

For cold molecule collisions, it is very important to have a good
representation of the long-range forces. These include a large
contribution from dispersion (intermolecular correlation), so
require a correlated treatment. We therefore use the MCSCF
orbitals in an MRCI calculation, again of the lowest three
electronic states. The MOLPRO package implements the ``internally
contracted" MRCI algorithm of Werner and Knowles \cite{Wer}. The
reference space in the MRCI is the same as the active space for
the MCSCF, and single and double excitations are included from all
orbitals except oxygen 1s. As described in Appendix 1, the two
$^1A'$ states are calculated in a single MRCI block, so that they
share a common basis set.

We encountered difficulties with non-degeneracy between the two
components of the $^1\Pi$ states at linear geometries. These are
described in Appendix 1. However, using the basis sets and
procedures described here, the non-degeneracies were never greater
than 90 $\mu E_h$ in the total energies for distances $R\ge 2.25$
\AA\ (and considerably less in the interaction energies around the
linear minimum).

\subsection{Transforming to a diabatic representation}

As described above, the two surfaces of $^1A'$ symmetry cross at
conical intersections at linear geometries. For dynamical
calculations, it is highly desirable to transform the adiabatic
states into diabatic states (or, strictly, quasidiabatic states).
MOLPRO contains code to carry out diabatization by maximizing
overlap between the diabatic states and those at a reference
geometry. However, this did not work for our application because
we were unable to find reference states that had enough overlap
with the lowest adiabats at all geometries. We therefore adopted a
different approach, based on matrix elements of angular momentum
operators. We use a Cartesian coordinate system with the $z$ axis
along the OH bond. At any linear geometry, the $\Pi$ component of
$^1A'$ symmetry is uncontaminated by the ion-pair state, and the
matrix elements of $\hat L_z$ are
\begin{eqnarray}
\langle ^1A'(\Pi)    | \hat L_z | ^1\App \rangle &=& i; \nonumber\\
\langle ^1A'(\Sigma) | \hat L_z | ^1\App \rangle &=& 0.
\end{eqnarray}
At nonlinear geometries, the actual $^1A'$ states can be
represented approximately as a mixture of $\Sigma$ and $\Pi$
components,
\begin{equation}
\left(\begin{matrix}\Psi_{1A'} & \Psi_{2A'}\end{matrix}\right) =
\left(\begin{matrix}\Psi_\Pi & \Psi_\Sigma\end{matrix}\right)
\left(\begin{matrix}\cos\phi & \sin\phi \\ -\sin\phi &
\cos\phi\end{matrix}\right), \label{mix}
\end{equation}
where the ``singlet" superscripts have been dropped to simplify
notation. If Eq.\ \ref{mix} were exact, the matrix elements of
$\hat L_z$ would be
\begin{eqnarray}
\langle 1A' | \hat L_z | \App \rangle &=& i\cos\phi; \nonumber\\
\langle 2A' | \hat L_z | \App \rangle &=& i\sin\phi.
\end{eqnarray}
The mixing angle $\phi$ would thus be given by
\begin{equation}
\phi = \tan^{-1} \frac{\langle 2A'| \hat L_z | \App \rangle}
{\langle 1A' | \hat L_z | \App \rangle}. \label{ratio}
\end{equation}
In the present work, we have taken the mixing angle to be {\it
defined} by Eq.\ \ref{ratio}, using matrix elements of $\hat L_z$
calculated between the MRCI wavefunctions. This gives a mixing
angle that, for linear geometries, is $\phi=0$ at long range and
$\phi=\pi/2$ at short range (inside the crossing).

One complication that arises here is that the signs of the three
wavefunctions are arbitrary, and may change discontinuously from one
geometry to another. The signs of the matrix elements obtained
numerically by MOLPRO are thus completely arbitrary. It was
therefore necessary to pick a sign convention for the matrix
elements at linear geometries and adjust the signs at other
geometries to give a smoothly varying mixing angle.

It should be noted that this diabatization procedure is not
general, and will fail if there is any geometry where both the
numerator and the denominator of Eq.\ \ref{ratio} are small.
Fortunately, this was not encountered for RbOH. The sum of squares
of the two matrix elements of $\hat L_z$ was never less than 0.99
at distances from $R=3.0$ \AA\ outwards, and never less than 0.7
even at $R=2.0$ \AA.

\begin{figure}
\begin{center}
\epsfig{file=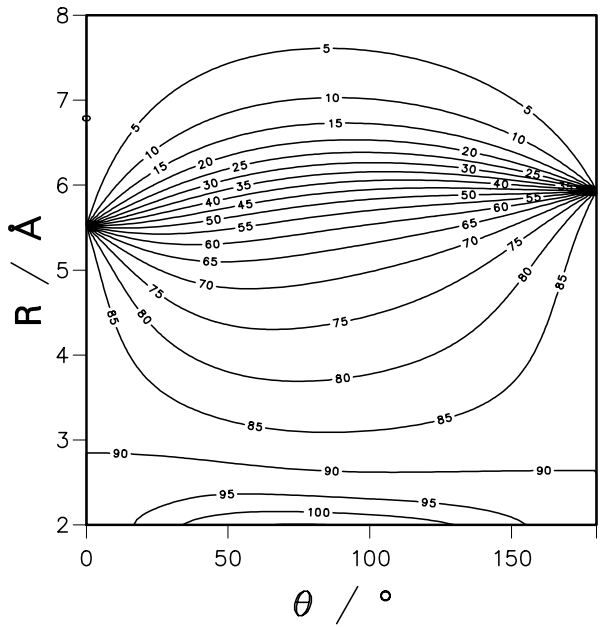,angle=-90,width=0.95\linewidth,clip=}
\end{center}
\caption{Contour plot of the diabatic mixing angle $\phi$ (in
degrees) for the $^1A'$ states of RbOH} \label{figmix}
\end{figure}

The mixing angles obtained for the singlet states of RbOH are
shown as a contour plot in Figure \ref{figmix}. As expected,
$\phi$ changes very suddenly from 0 to $90^\circ$ at linear and
near-linear geometries, but smoothly at strongly bent geometries.

Once a smooth mixing angle has been determined, the diabatic
potentials and coupling surface are obtained from
\begin{eqnarray}
\left(\begin{matrix}H_{11} & H_{12} \\ H_{21} & H_{22}\end{matrix}\right) &=&
\left(\begin{matrix}\cos\phi & \sin\phi \cr -\sin\phi & \cos\phi\end{matrix}\right)
\nonumber\\ &\times& \left(\begin{matrix}E_{1A'} & 0 \\ 0 &
E_{2A'}\end{matrix}\right) \left(\begin{matrix}\cos\phi & -\sin\phi \\ \sin\phi &
\cos\phi\end{matrix}\right).
\end{eqnarray}

\begin{figure}
\setlength{\unitlength}{4mm}
\begin{picture}(0,0)(0,0)
\put(7.5,-3.5){\makebox(0,0){{\bf \large{(a)}}}}
\put(7.5,-25){\makebox(0,0){{\bf \large{(b)}}}}
\end{picture}
\vspace{-.6cm}
\begin{center}
\epsfig{file=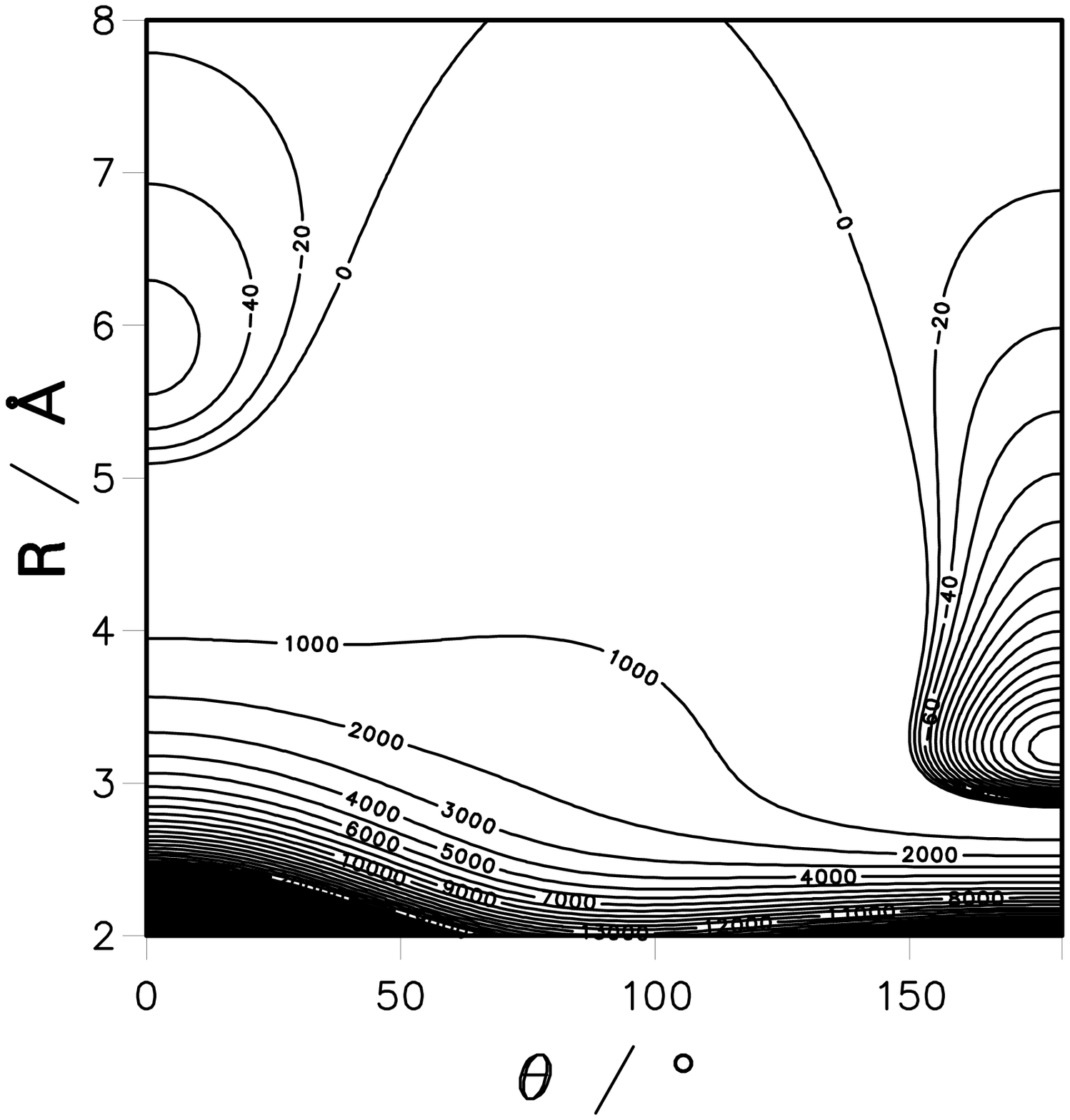,angle=-90,width=0.95\linewidth,clip=}
\epsfig{file=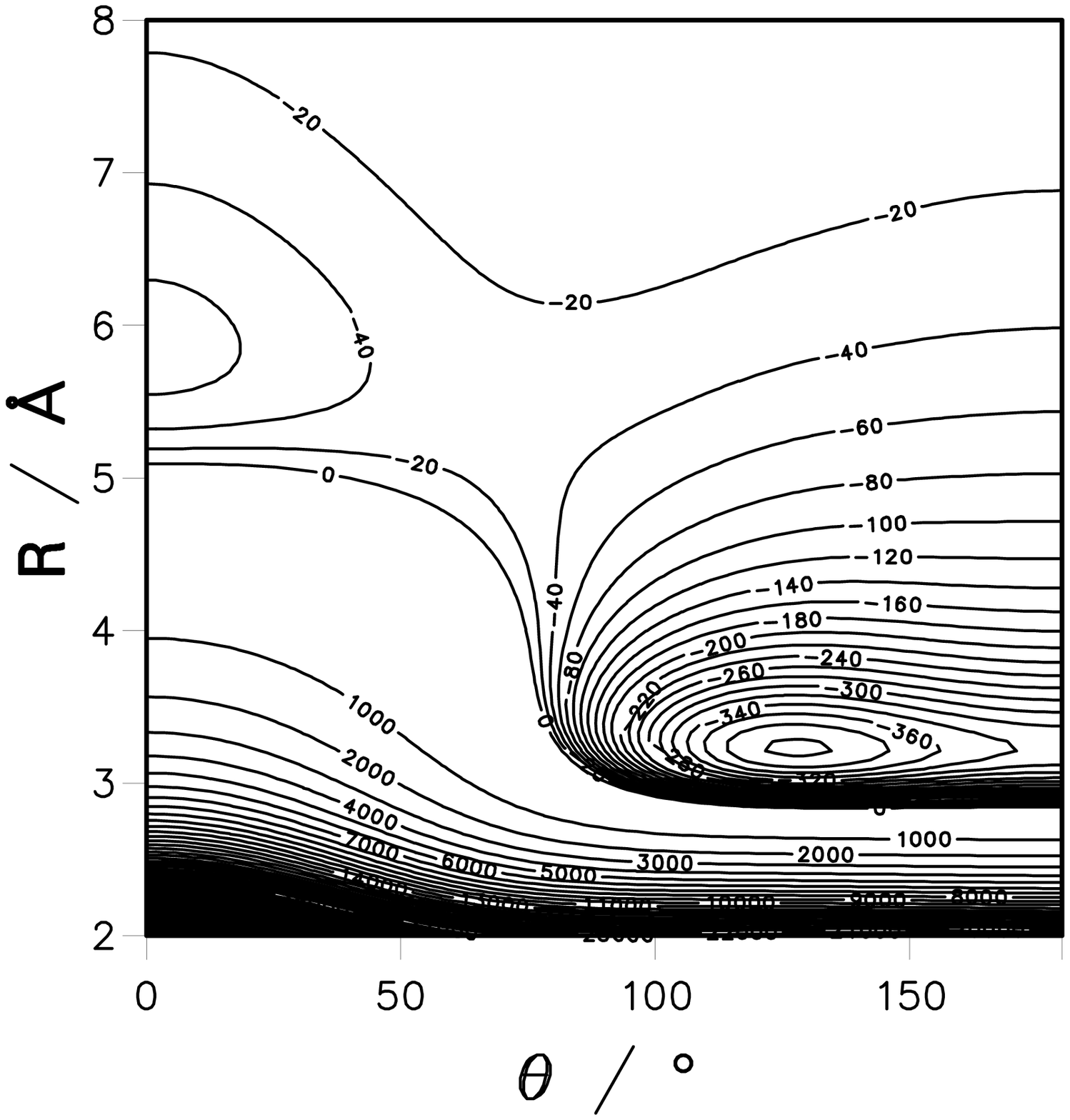,angle=-90,width=0.95\linewidth,clip=}
\end{center}
\caption{Contour plot of the diabatic $^1A'$ (panel (a)) and
$^1\App$ (panel (b)) covalent (non-ion-pair) potential energy
surfaces for RbOH from MRCI calculations. Contours are labeled in
cm$^{-1}$.} \label{figsing1}
\end{figure}
\begin{figure}
\setlength{\unitlength}{4mm}
\begin{picture}(0,0)(0,0)
\put(7.5,-3.5){\makebox(0,0){{\bf \large{(a)}}}}
\put(7.5,-25){\makebox(0,0){{\bf \large{(b)}}}}
\end{picture}
\vspace{-.5cm}
\begin{center}
\epsfig{file=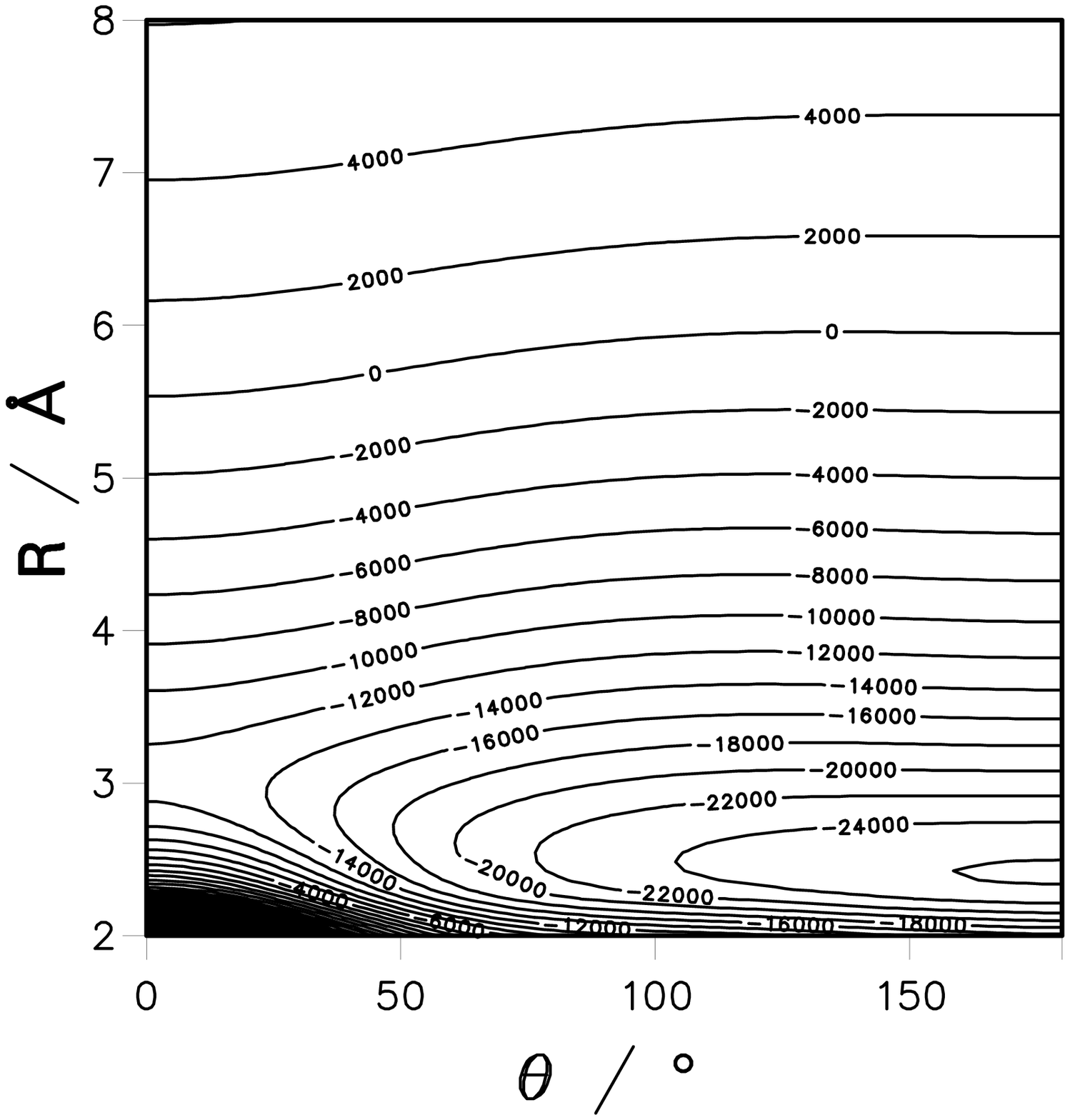,angle=-90,width=0.95\linewidth,clip=}
\epsfig{file=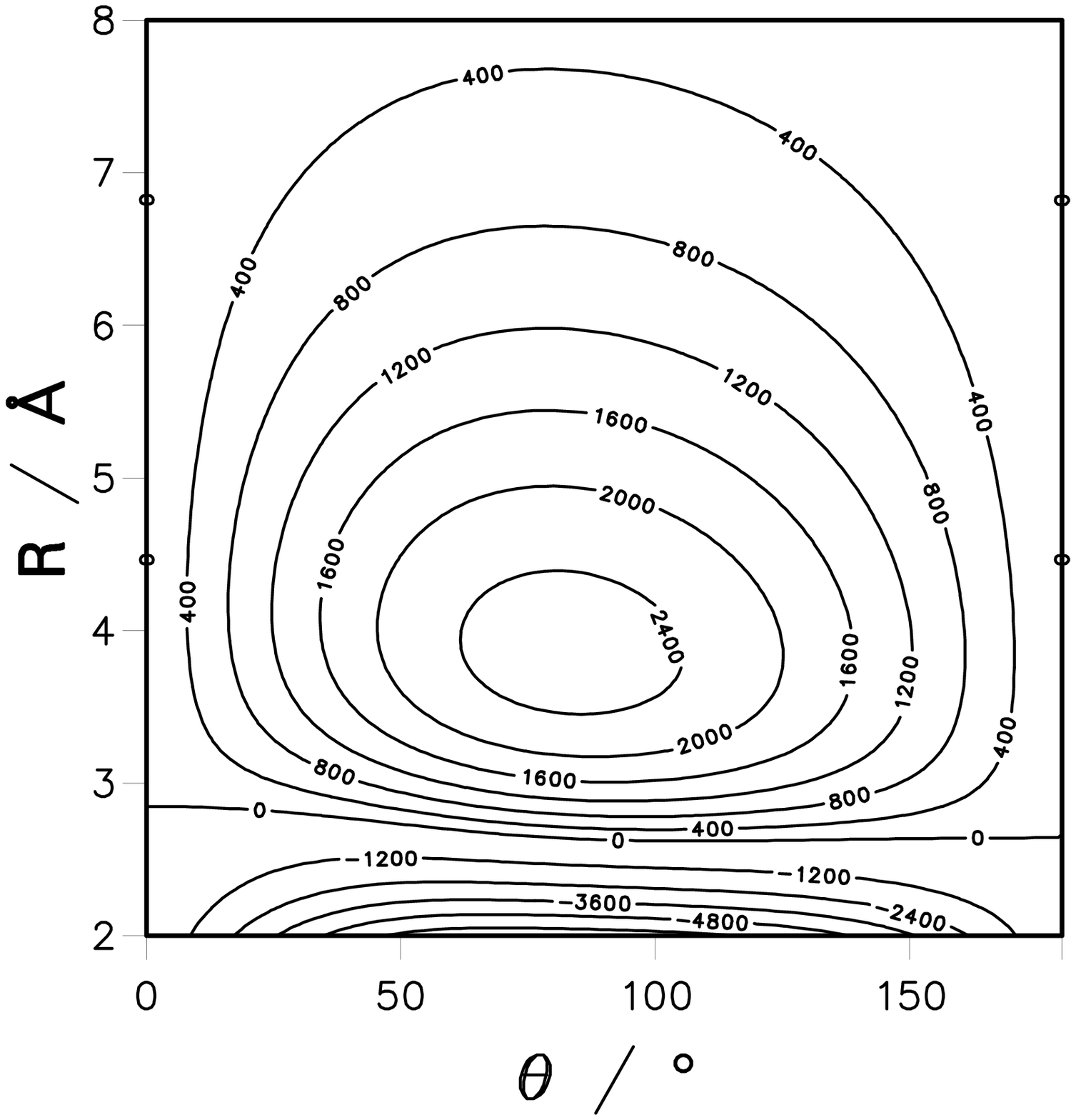,angle=-90,width=0.947\linewidth,clip=}
\end{center}
\caption{Contour plot of the $^1A'$ diabatic ion-pair potential
energy surface (panel (a)) and the diabatic coupling potential
(panel (b)) for RbOH from MRCI calculations. Contours are
labeled in cm$^{-1}$.} \label{figsing2}
\end{figure}
The diabatization was carried out using total electronic energies
(not interaction energies). The two singlet diabatic PES that correlate
with Rb($^2S$) + OH($^2\Pi$) are shown in Figure \ref{figsing1},
and the diabatic ion-pair surface and the coupling potential are
shown in Figure \ref{figsing2}. Some salient characteristics of
the surfaces are given in Table \ref{tabpot}. As required, the
$^1\App$ and $^1A'$ covalent states are degenerate at the two
linear geometries, with a relatively deep well (337 cm$^{-1}$) at
Rb-OH and a much shallower well at Rb-HO. At bent geometries, it
is notable that the potential well is broader and considerably
deeper for the $^1\App$ state than for the $^1A'$ state; indeed,
the $^1A'$ state has a linear minimum, while the $^1\App$ has a
bent minimum at $\theta=128^\circ$ with a well depth of 405
cm$^{-1}$. In spectroscopic terms, this corresponds to a
Renner-Teller effect of type 1(b) in the classification of Pople
and Longuet-Higgins \cite{Pop58}.

\begin{table}
\caption{Characteristics of the diabatic PES
for RbOH.} \vspace{4mm}
\begin{tabular}{lcccccc}
\hline\hline & \ & $^3\App$ & $^3A'$ & $^1\App$
& $^1A'$ & $^1A'$ \\
&&&&& covalent & ion-pair \\
\hline
well depth/cm$^{-1}$      &&   615 &   511 &   405 &   337 & 26260 \\
distance of minimum/\AA   && 3.185 & 3.170 & 3.226 & 3.230 & 2.407 \\
angle of minimum/ $^\circ$ &&   123 &   180 &   128 &   180 &   180 \\
\hline
\end{tabular}
\label{tabpot}
\end{table}

The fact that the $^1\App$ state is deeper than the $^1A'$ state
is somewhat unexpected, and will be discussed in more detail in
the context of the triplet surfaces below. The fact that the
surface for the covalent $^1A^\prime$ state is slightly repulsive
between $R=4$ and 9 \AA\ and between $\theta=40^\circ$ and
$150^\circ$ is an artifact of the diabatization procedure and
should not be given physical significance: the choice of mixing
angle (\ref{ratio}) is an approximation, and a slightly different
choice would give different diabats and coupling terms, but of
course corresponding to the same adiabatic surfaces.

The ion-pair state has a deep well at the RbOH geometry and a rather
shallower one at RbHO, as expected from electrostatic
considerations. The region around the minimum of this surface has
been characterized in more detail by Lee and Wright \cite{Lee03}.

It is notable that the coupling potential is quite large, peaking
at 2578 cm$^{-1}$ at $R=3.87$ \AA\ and $\theta=83^\circ$, and is
thus larger than the interaction energy for the two covalent
surfaces at most geometries. It may be seen from Figure
\ref{figmix} that the $\Sigma-\Pi$ mixing it induces is
significant at most distances less than 8 \AA.

\subsection{Triplet states}

The triplet states of RbOH are considerably simpler than the singlet
states, because there is no low-lying triplet ion-pair state. There
is thus no conical intersection, and diabatization is not needed.

Single-reference calculations would in principle be adequate for
the triplet states. Nevertheless, for consistency with the singlet
surfaces, we carried out MCSCF and MRCI for the triplet surfaces
as well. The (10,3) active and reference spaces were specified as
for the singlet states, except that there is only one relevant
$^3A'$ state and the state average in the MCSCF calculation is
therefore over the lowest two states.

\begin{figure}
\setlength{\unitlength}{4mm}
\begin{picture}(0,0)(0,0)
\put(7.5,-3.5){\makebox(0,0){{\bf \large{ (a)}}}}
\put(7.5,-25){\makebox(0,0){{\bf \large{(b)}}}}
\end{picture}
\vspace{-.5cm}
\begin{center}
\epsfig{file=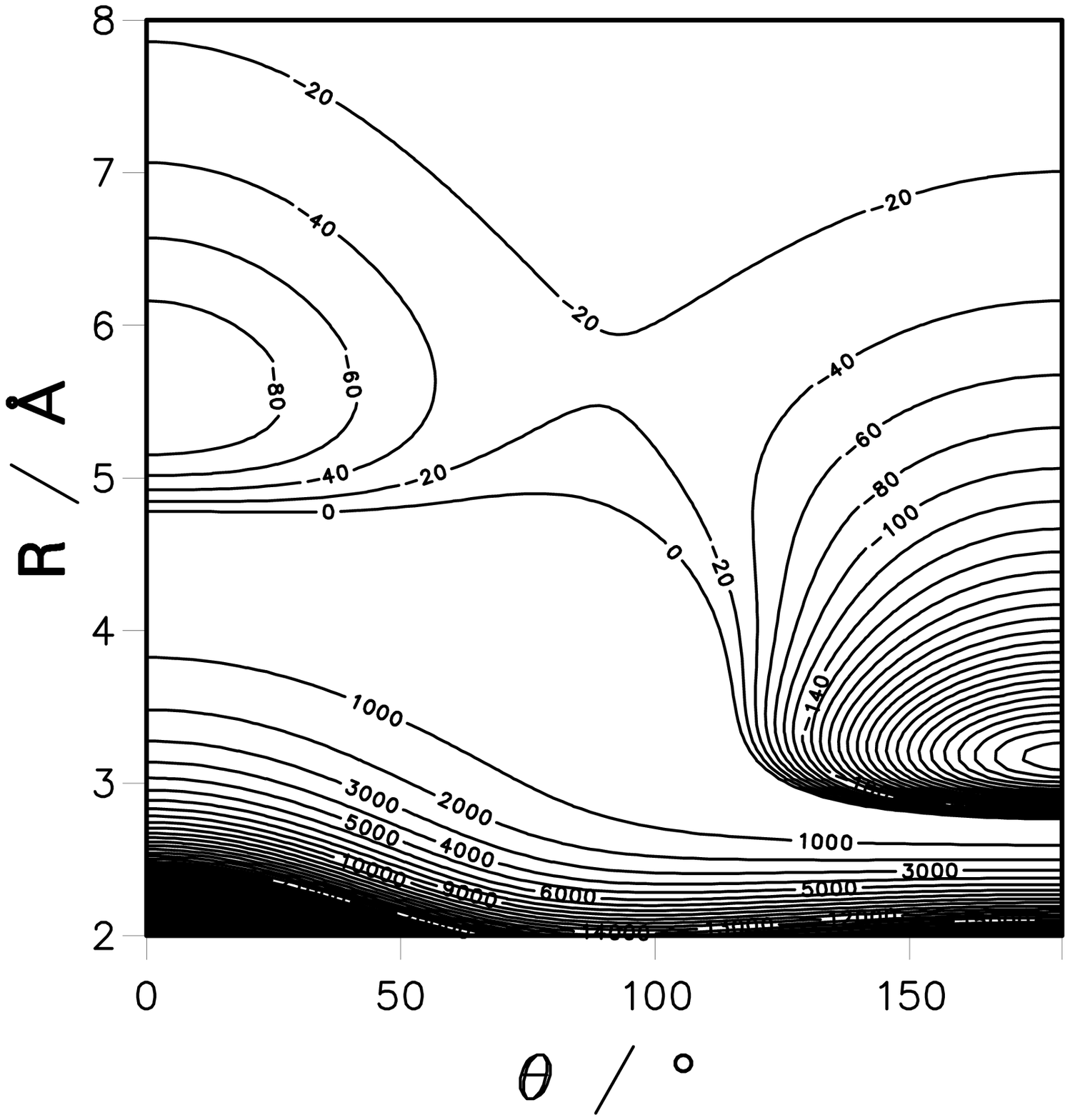,angle=-90,width=0.95\linewidth,clip=}
\epsfig{file=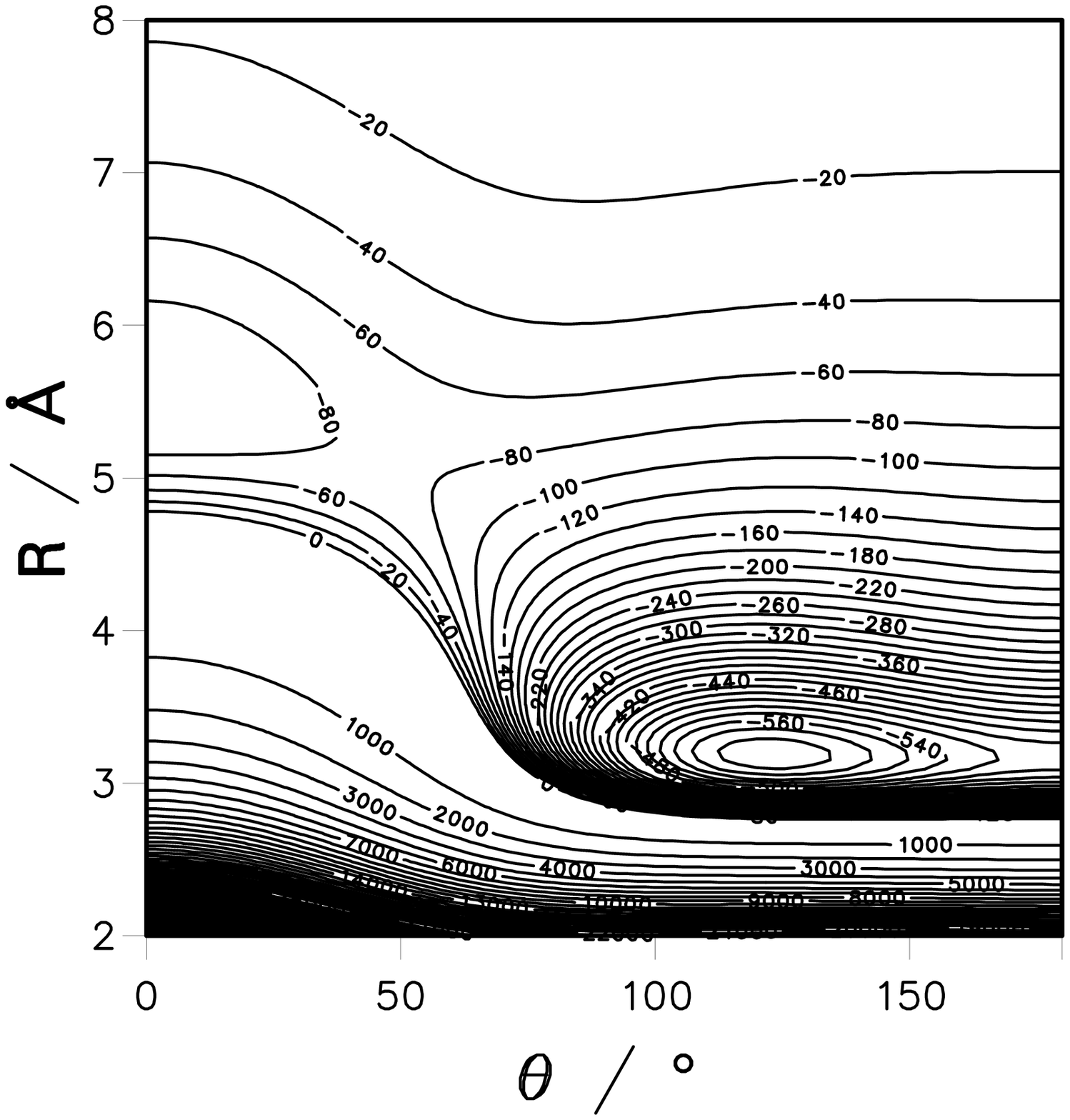,angle=-90,width=0.95\linewidth,clip=}
\end{center}
\caption{Contour plot of the $^3A'$ (panel (a)) and $^3\App$
(panel (b)) PES for RbOH from MRCI
calculations. Contours are labeled in cm$^{-1}$.} \label{figtrip}
\end{figure}

Contour plots of the $^3A'$ and $^3\App$ surfaces are shown in
Figure \ref{figtrip}. As for the singlet states, it is notable
that the $^3\App$ state lies below the $^3A'$ state at nonlinear
geometries. This ordering is different from that found for systems
such as Ar-OH \cite{Esp90,H93ArOH} and He-OH \cite{Lee00}. In each
case, the $A'$ state corresponds to an atom approaching OH {\it
in} the plane of the unpaired electron, while the $\App$ state
corresponds to an atom approaching OH {\it out of} the plane. For
He-OH and Ar-OH, the $^2A'$ state is deeper than the $^2\App$
state simply because there is slightly less repulsion due to a
half-filled $\pi$ orbital than due to a doubly filled $\pi$
orbital. Since these systems are dispersion-bound, and the
dispersion coefficients are similar for the $^2A'$ and $^2\App$
states, the slightly reduced repulsion for the $^2A'$ state
produces a larger well depth.

Rb-OH is quite different. The long-range coefficients still
provide a large part of the binding energy of the covalent states,
but the equilibrium distances (around 3.2 \AA, Table \ref{tabpot})
are about 1 \AA\ shorter than would be expected from the sum of
the Van der Waals radii of Rb (2.44 \AA) and OH (1.78 \AA,
obtained from the Ar value of 1.88 \AA\ and the Ar-OH equilibrium
distance of 3.67 \AA\ \cite{H93ArOH}).

\begin{figure}
\begin{center}
\epsfig{file=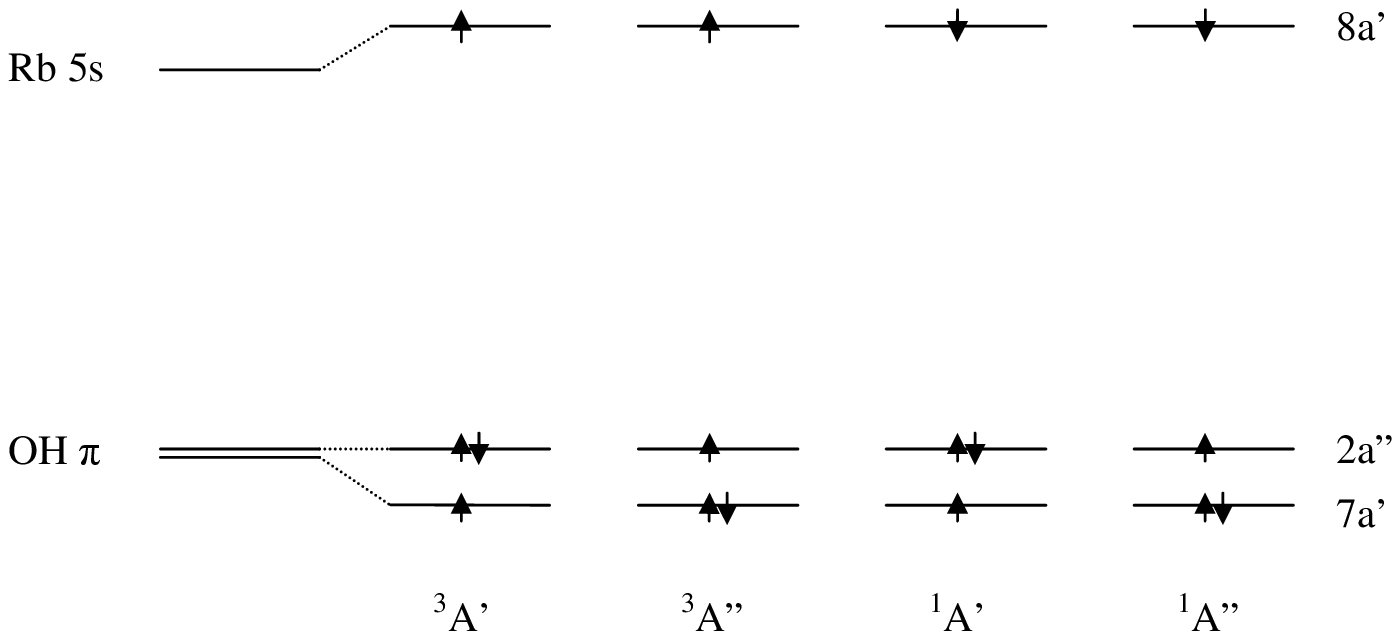,angle=0,width=0.95\linewidth,clip=}
\end{center}
\caption{Orbital occupations for the covalent (non-ion-pair)
states of RbOH at nonlinear geometries, $90^\circ < \theta <
180^\circ$, showing the origin of the reduced repulsion for $\App$
states compared to $A'$ states.} \label{figbond}
\end{figure}

The qualitative explanation is that at nonlinear geometries there
is significant overlap between the Rb 5s orbital and the OH $\pi$
orbital of $a'$ symmetry, forming weakly bonding and antibonding
molecular orbitals (MOs) for the RbOH supermolecule. The effect of
this is shown in Figure \ref{figbond}. In the $A'$ states, the
bonding and antibonding MOs are equally populated and there is no
overall stabilization. However, for the $\App$ states the bonding
MO is doubly occupied and the antibonding MO is singly occupied.
This gives a significant reduction in the repulsion compared to a
simple overlap-based model.

Even at linear Rb-OH geometries, the repulsion is reduced by a
similar effect involving the Rb 5s orbital and the highest
occupied OH $\sigma$ orbital ($6a'$ in Figure \ref{figorb}).

\subsection{Converting MRCI total energies to interaction energies}

The MRCI procedure is not size-extensive, so cannot be corrected
for basis-set superposition error (BSSE) using the counterpoise
approach \cite{Boy}. In addition, there are Rb 5p orbitals that
lie 1.87 eV above the ground state. This is below the ion-pair
energy at large $R$, so that a (10,3) active space includes
different orbitals asymptotically and at short range. It is thus
hard to calculate asymptotic energies that are consistent with the
short-range energies directly. Nevertheless, for collision
calculations we ultimately need interaction energies, relative to
the energy of free Rb($^2S$) + OH($^2\Pi$).

To circumvent this problem, we obtained angle-dependent long-range
coefficients $C_6(\theta)$ and $C_7(\theta)$ for the {\it triplet}
states of RbOH and used these to extrapolate from 12\,\AA\ outwards
at each angle. We carried out RCCSD calculations (restricted
coupled-cluster with single and double excitations) on the $^3A'$
and $^3\App$ states at all angles at distances of 15, 25 and
100\,\AA. RCCSD was chosen in preference to RCCSD(T) for greater
consistency with the MRCI calculations. Coupled cluster calculations
are size-extensive, so in this case the interaction energies
$V(R,\theta)$ for the $^3A'$ and $^3\App$ states were calculated
including the counterpoise correction \cite{Boy}. The interaction
energy at 100\,\AA\ was found to be non-zero (about $-1.38\,\mu
E_h$), but was the same to within $10^{-9}\,E_h$ for both $^3A'$ and
$^3\App$ states and at all angles. For each pair of angles $\theta$
and $\pi-\theta$, the energies were fitted to the form
\begin{equation}
E(R,\theta) = E_\infty -C_6(\theta) R^{-6} -C_7(\theta)R^{-7}
-C_8(\theta) R^{-8}, \label{c678}
\end{equation}
with the constraints that
\begin{eqnarray}
C_6(\pi-\theta)&=&C_6(\theta) \\ C_7(\pi-\theta)&=&-C_7(\theta).
\end{eqnarray}
Sums and differences of the long-range coefficients for the two
states,
\begin{eqnarray}
C_n^0(\theta)&=&\frac{1}{2}\left[C_n^{\App}(\theta)+C_n^{A'}(\theta)\right]\\
C_n^2(\theta)&=&\frac{1}{2}\left[C_n^{\App}(\theta)-C_n^{A'}(\theta)\right],
\end{eqnarray}
were then smoothed by fitting to the theoretical functional forms
\begin{eqnarray}
C_6^0(\theta)&=&C_6^{00} + C_6^{20} P_2^0(\cos\theta) \\
C_6^2(\theta)&=&C_6^{22} P_2^2(\cos\theta) \\
C_7^0(\theta)&=&C_7^{10} P_1^0(\cos\theta) + C_7^{30} P_3^0(\cos\theta) \\
C_7^2(\theta)&=&C_7^{32} P_3^2(\cos\theta),
\end{eqnarray}
where $P_\lambda^\nu(\cos\theta)$ are associated Legendre
functions. The coefficients obtained by this procedure are
summarized in Table \ref{tabdisp}. The resulting smoothed values
of $C_6^0(\theta)$, $C_6^2(\theta)$, $C_7^0(\theta)$ and
$C_7^2(\theta)$ were then used to generate $C_6^{A'}(\theta)$,
$C_6^{\App}(\theta)$, $C_7^{A'}(\theta)$ and $C_7^{\App}(\theta)$.
Finally, the MRCI {\it total} energies at $R=10$ and 12\,\AA\ for
both singlet and triplet states were refitted to Eq.\ \ref{c678},
with $C_6(\theta)$ and $C_7(\theta)$ held constant at the smoothed
values, to obtain an MRCI value of $E_\infty$ for each angle (and
surface). These angle-dependent values of $E_\infty$ were used to
convert the MRCI total energies into interaction energies.

It should be noted that the long-range coefficients in Rb-OH have
substantial contributions from induction as well as dispersion. A
simple dipole-induced dipole model gives $C_{\rm6,ind}^{00} =
C_{\rm6,ind}^{20} = 133\ E_ha_0^6$ and accounts for 40\% of
$C_{\rm6}^{00}$ and 90\% of $C_{\rm6}^{20}$

\begin{table}
\caption{Long-range coefficients for Rb-OH obtained by fitting to
RCCSD calculations on triplet states.} \vspace{4mm}
\begin{tabular}{ccccc}
\hline\hline $\lambda$& 0 & 1 & 2 & 3 \\
\hline
$C_6^{\lambda0}/E_ha_0^6$ & 325.0 & -- & 151.0 & -- \\
$C_6^{\lambda2}/E_ha_0^6$ & -- & -- & 1.9 & -- \\
$C_7^{\lambda0}/E_ha_0^7$ & -- & 1035.4 & -- & 630.0 \\
$C_7^{\lambda2}/E_ha_0^7$ & -- & -- & -- & --40.1 \\
\hline
\end{tabular}
\label{tabdisp}
\end{table}

\subsection{Interpolation and fitting}
The procedures described above produce six potential energy
surfaces on a two-dimensional grid of geometries $(R,\theta)$:
four surfaces corresponding to covalent states (``c''), with $A'$
or $A''$ symmetry and triplet or singlet multiplicity, one surface
with ion-pair character with $^1A'$ symmetry, and finally the
non-vanishing coupling of this latter state with the covalent
$^1A'$ state. We will denote the 6 surfaces $V^{^1A'_{\rm c}}$,
$V^{^1A''_{\rm c}}$, $V^{^3A'_{\rm c}}$, $V^{^3A''_{\rm c}}$,
$V^{^1A'_{\rm i}}$, and $V^{\rm ic}$, respectively. Labels will be
dropped below when not relevant to the discussion.

For the covalent states of each spin multiplicity, interpolation
was carried out on sum and difference potentials,
\begin{eqnarray}
V_0(R,\theta)&=&\frac{1}{2}\left[V^{\App}(R,\theta)+V^{A'}(R,\theta)\right]\\
V_2(R,\theta)&=&\frac{1}{2}\left[V^{\App}(R,\theta)-V^{A'}(R,\theta)\right],
\end{eqnarray}
with the difference potentials set to zero at $\theta=0$ and
$180^\circ$ to suppress the slight non-degeneracy in the MRCI
results. Our approach to two-dimensional interpolation follows
that of Meuwly and Hutson \cite{H99NeHFmorph} and Sold\'an {\em et
al.}\ \cite{Soljcp02a}. The interpolation was carried out first in
$R$ (for each surface and angular point) and then in $\theta$. The
interpolation in $R$ used the RP-RKHS (reciprocal power
reproducing kernel Hilbert space) procedure \cite{Ho96} with
parameters $m=5$ and $n=3$. This gives a potential with long-range
form
\begin{equation}
V_q(R,\theta) = -C_6^q(\theta) R^{-6} -C_7^q(\theta)R^{-7}
-C_8^q(\theta) R^{-8}.
\end{equation}
The values of $C_6^q(\theta)$ and $C_7^q(\theta)$ were fixed
at the values described in the previous subsection
 \cite{Ho00,Sol00}.

For the ion-pair state, it is the quantity $V^{\rm i}(R,\theta) -
E_\infty^{\rm i}$ that is asymptotically zero, where
$E_\infty^{\rm i}=0.0863426\,E_h$ is the ion-pair threshold. This
was interpolated using RP-RKHS interpolation with parameters $m=0$
and $n=2$, which gives a potential with long-range form
\begin{equation}
V^{\rm i}(R,\theta) = E_\infty^{\rm i} - C_1(\theta)R^{-1}
-C_2(\theta) R^{-2}.
\end{equation}
The coefficient $C_1(\theta)$ was fixed at the Coulomb value of
$1\,E_ha_0^{-1}$.

The coupling potential $V^{\rm ic}(R,\theta)$ has no obvious
inverse-power form at long range. It was therefore interpolated
using the ED-RKHS (exponentially decaying RKHS) approach
 \cite{Hol99}, with $n=2$ and $\beta=0.77$\,\AA. This gives a
potential with long-range form
\begin{equation}
V^{\rm ic}(R,\theta) = A(\theta) \exp(-\beta R).
\end{equation}
The value of $\beta$ was chosen by fitting the values of the
coupling potential at $R=10$ and 12\,\AA\ to decaying exponentials.

Interpolation in $\theta$ was carried out in a subsequent step.
An appropriate angular form is
\begin{equation}
V_q(R,\theta) = \sum_k V_{kq}(R) \Theta_{kq}(\cos\theta),
\label{fit}
\end{equation}
where $\Theta_{kq}(\cos\theta)$ are normalized associated
Legendre functions, \begin{equation}
\Theta_{kq}(\cos\theta)=
\sqrt{\left(\frac{2k+1}{2}\right)
\frac{(k-|q|)!}{(k+|q|)!}}
P_k^{|q|}(\cos\theta),
\end{equation}
and $q=0$ for the sum and ion-pair potentials, 2 for the
difference potentials and 1 for the coupling potential. $q$ can
thus be used as a label to distinguish the different potentials.
The coefficients $V_{kq}(R)$ for $k=q$ to 9 were projected out
using Gauss-Lobatto quadrature, with weights $w_i$,
\begin{equation}
V_{kq}(R) = \sum_i w_i V_q(R,\theta_i)
\Theta_{kq}(\cos\theta_i),
\end{equation}
Since there are fewer coefficients than points, the resulting
potential function does not pass exactly through the potential
points. However, the error for the covalent states in the well
region is no more than 20 $\mu E_h$.

We thus arrive at a set of $R$-dependent coefficients
$V_{kq}^{\alpha,2S+1}(R)$, with $\alpha={\rm c}$, i or ic
labeling the potentials for the pure covalent or ion-pair states
or the coupling between them and $2S+1=0$ or 1 for singlet or
triplet states respectively. These coefficients will be used below
in evaluating the electronic potential matrix elements which
couple collision channels in the dynamical calculations.

\vspace{-.2cm}
\section{Dynamical methodology}
\vspace{-.2cm}
\subsection{The basis sets}
We carry out coupled-channel calculations of the collision
dynamics. The channels are labeled by quantum numbers that
characterize the internal states of the colliding partners, plus
partial wave quantum numbers that define the way the partners
approach each other.

It is convenient to distinguish between the laboratory frame,
whose $z$ axis is taken to be along the direction of the external
field (if any) and the molecule frame, whose $z$ axis lies along
the internuclear axis of the OH molecule and whose $xz$ plane
contains the triatomic system. This allows us to define {\em
external coordinates} that fix the collision plane, and {\em
internal coordinates} that describe the relative position of the
components on it. As external coordinates, we choose the Euler
angles $(\alpha,\beta,\gamma)$ required to change from the
laboratory frame to the molecule-frame; as internal coordinates we
use the system of Jacobi coordinates ($R,\theta$) defined above. We also
define the spherical angles ($\theta', \phi'$) that describe the
orientation of the intermolecular axis in the external
(laboratory) frame.

We first focus on the covalent channels OH$(\pi^3,\ ^2\Pi)+{\rm
Rb}(5s^1,\ ^2S_{1/2})$. The OH molecule can be described using
Hund's case (a) quantum numbers: the internal state is expressed
in a basis set given by $ |s_d \sigma \rangle | \lambda \rangle |j
m \omega \rangle $, where $s_d$ is the electronic spin of OH and
$\sigma$ its projection on the internuclear axis; $\lambda$ is the
projection of the electronic orbital angular momentum onto the
internuclear axis; $j$ is the angular momentum resulting from the
electronic and rotational degrees of freedom, $m$ its projection
on the laboratory axis and $\omega=\lambda+\sigma$ its projection
on the internuclear axis. The symmetric top wavefunctions that
describe the rotation of the diatom in space are defined by $ |j m
\omega \rangle = \sqrt{\frac{2j+1}{8 \pi^2}} D_{ m \omega
}^{j*}(\alpha,\beta,\gamma) $. At this stage $\lambda$, $\sigma$
and $\omega$ are still signed quantities. However, in zero
electric field, energy eigenstates of OH are also eigenstates of
parity, labeled $\epsilon=e$ or $f$. These labels refer to
``+``or ``$-$'' sign taken in the combination of $\omega$ and
$-\omega$; the real parity is given by $\epsilon (-1)^{j-s_d}$. To
include parity, we can define states $ |s_d \bar{\lambda}
\bar{\omega} \epsilon j m \rangle $, where the bar indicates the
absolute values of a signed quantity. Finally, we need to include
also the nuclear spin degree of freedom: if $i_d$ designates the
nuclear spin angular momentum of the diatom, then $j$ and $i_d$
combine to form $f_d$, the total angular momentum of the diatom,
which has projection $m_{f_d}$ on the laboratory $z$ axis. The
resulting basis set that describes the physical states of the OH
molecule is
\begin{eqnarray} |s_d \bar{\lambda} \bar{\omega} \epsilon (j i_d )f_d
m_{f_d}. \rangle \end{eqnarray} For Rb, the electronic angular
momentum is given entirely by the spin $s_a$ of the open-shell
electron, which combines with the nuclear spin $i_a$ to form
$f_a$, the total angular momentum of the atom. The state of the Rb
atom can then be expressed as
\begin{eqnarray}
|(s_a i_a )f_a m_{f_a} \rangle.
\end{eqnarray}
The explicit inclusion of the $s_d$ and $s_a$ quantum numbers
allows us to use the same notation for ion-pair channels
OH$^-(\pi^4,\ ^1\Sigma^+ )+ {\rm Rb}^+ (^1S_0)$: in this case both
partners are closed-shell, so
$s_d=s_a=\bar{\lambda}=\bar{\omega}=0$ and $\epsilon=+1$.

The resulting basis set for close-coupling calculations on the
complete system (designated B1) has the form
\begin{eqnarray} | B1 \rangle = |s_d \bar{\lambda} \bar{\omega} \epsilon (j i_d )f_d m_{f_d}
 \rangle |(s_a i_a )f_a m_{f_a} \rangle | L M_L \rangle, \end{eqnarray}
where $| L M_L \rangle $ denotes the partial wave degree of
freedom and is a function of the $(\theta',\phi')$ coordinates
considered above. The ket $ | L M_L \rangle$ corresponds to
$Y_{LM_L}(\theta',\phi')$. Ultimately, S-matrix elements for
scattering are expressed in basis set B1.

The scattering Hamiltonian is block-diagonal in {\em total angular
momentum} and {\em total parity}.
The total parity is well defined in the basis set we have
selected, and given by $p=\epsilon (-1)^{j-s_d+L}$. It is
conserved in the presence of a magnetic field but not an electric
field. The total angular momentum is not conserved in the presence
of either a magnetic or an electric field. However, the projection
of total angular momentum on the laboratory $z$ axis, given by
$M_{\mathcal J}=m_{f_d}+m_{f_a}+M_L$, is conserved in the presence
of an external field aligned with the laboratory $z$ axis.
\vspace{-.2cm}
\subsection{Matrix elements of the potential energy}

The PES in section II are diagonal in the
total electronic spin, $S$, and in the states of the nuclei,
$m_{i_a}$ and $m_{i_d}$, with $m_{i_a}$ and $ m_{i_d}$ the nuclear spin
projections of the atom and diatom in the laboratory frame.
 We therefore find it convenient to define
basis sets labeled by these quantum numbers. This allows not
only the direct calculation of potential matrix elements, but the
definition of some useful {\em frame transformations} \cite{Fano}.
  Two other basis
sets, B2 and B3, defined with/without parity (B2p/B2w and B3p/B3w)
are described in Appendix 2.
The corresponding frame transformations are defined in section
\ref{Solve}.

The calculation of the matrix elements of the potential energy is
most direct in basis set B2w. This is based on Hund's case (b)
quantum numbers for the molecule, and is given by
\begin{eqnarray} | B2w \rangle = | (s_d s_a) S M_S \rangle
| n m_n \lambda \rangle | i_d m_{i_d} \rangle | i_a m_{i_a}
\rangle | L M_L \rangle,
\end{eqnarray}
where $n$ is the total angular momentum excluding spin, with
projection $\lambda$ on the internuclear axis, $|S M_S \rangle$
indicates the total spin state of the electrons, and $ | i_a
m_{i_a} \rangle | i_d m_{i_d} \rangle $ indicates the states of
the nuclei.

In order to relate the PES obtained in
section II to the quantum numbers of our channels, it is
convenient to recast the electronic wavefunctions for the covalent
states of $^{2S+1} A'$ and $^{2S+1} A''$ in terms of functions
with definite values of $\lambda$,
 \begin{eqnarray}
 |^{1(3)}{+1}_{\rm c}\rangle&& = -\frac{1}{\sqrt{2}}( |^{1(3)} A'_{\rm c}\rangle + i |^{1(3)} A''_{\rm c}\rangle); \nonumber \\
|^{1(3)} {-1}_{\rm c}\rangle&& = \frac{1}{\sqrt{2}}( |^{1(3)}
A'_{\rm c}\rangle - i |^{1(3)} A''_{\rm c}\rangle).
\end{eqnarray}
We can also associate $\lambda=0$ with the ion-pair wavefunction
(and denote it $ | ^{1}0 _{\rm i} \rangle $). Then the multipole
index $q$ in the potential expansion (\ref{fit}) is viewed as an
angular momentum transfer $q=\lambda' - \lambda $.

The B2w basis functions do not explicitly depend on $\theta$. We
therefore rotate the $\Theta_{kq}$ functions \cite{Brink} onto the
laboratory frame, to which molecular spins and partial waves are
ultimately referred. The functions $\Theta_{kq}$ are proportional
to renormalized spherical harmonics $C_{kq}(\theta,0)$
\cite{Brink}, for which
 \begin{eqnarray}
 C_{kq}(\theta,0)=\sum_\nu D_{\nu q }^k(\alpha,\beta,\gamma)
 C_{k\nu }(\theta', \phi').
 \end{eqnarray}
The potential now depends on the same angular coordinates as the
B2w basis functions. Integrating and applying the usual
relationships, we obtain the matrix elements of the electronic
potential in basis set $ | B2w \rangle $,
\begin{widetext}
\begin{eqnarray} \label{potenenes}
\lefteqn{\langle L M_L| \langle(s_d s_a) S M_S | \langle  n m_n \lambda
| V | n' m'_n \lambda' \rangle | (s'_d s'_a) S' M'_S \rangle| L'
M'_L \rangle \nonumber} \hspace{2cm} \\ &&= \delta_{S S'} \delta_{M_S M'_S}
\sum_k \sqrt{[n ] [n' ][L ][L' ]}(-1)^{(M_L+m'_n- \lambda')}
\left(\begin{array}{ccc} L & k &
 L' \\ -M_L & \nu & M_L' \end{array}\right)\left(\begin{array}{ccc} L & k &
 L' \\ 0 & 0 & 0 \end{array}\right)
 \nonumber
\\ &&\times \left(\begin{array}{ccc} n & k &
 n' \\ m_n & \nu & -m_n' \end{array}\right)
\left(\begin{array}{ccc} n & k &
 n' \\ \lambda & q & -\lambda' \end{array}\right)\sqrt{\frac{(2k+1)}{2}}
 \kappa V_{k |q|}^{\alpha,2S+1}(R),
\end{eqnarray}
\end{widetext}
where $[A]=2 A+1$, $q=\lambda' - \lambda$ and $\nu=m'_n-m_n$.
$\kappa$ is a constant whose value in the case of
covalent-covalent or ionic-ionic matrix elements ($q=0, \pm 2$) is
1. For covalent-ionic or ionic-covalent matrix elements ($q = \pm
1$), $\kappa = -\sqrt{1/2}$ or $\kappa = \sqrt{1/2}$ respectively.
In these last two cases, $V_{k |q|}^{\alpha,2S+1}(R)=V_{k1}^{\rm
ic,1}(R)$.

Our aim is to evaluate the matrix elements of the potential in
basis set B1. Basis set B3w, defined in Appendix 2, can be
considered as an intermediate step between B2w and B1. Starting
from Eq.\ \ref{potenenes} and changing basis to B3w (see Appendix
2), we obtain
\begin{widetext}
\begin{eqnarray}\label{elements}
\lefteqn{ \langle L M_L| \langle i_a m_{i_{a}} | \langle s_a m_{s_a} |
\langle i_d m_{i_{d}} | \langle j m \omega | \langle \lambda |
\langle s_d \sigma | V |s_d' \sigma' \rangle | \lambda' \rangle|j'
m' \omega' \rangle | i'_d m'_{i_{d}} \rangle |s_a' m'_{s_a}
\rangle | i'_a m'_{i_{a}} \rangle | L' M_L'\rangle } \nonumber\hspace{2cm}
\\ &=& \delta_{i_a i'_a} \delta_{m_{i_{a}} m'_{i_{a}} }
\delta_{i_d i'_d} \delta_{m_{i_{d}} m'_{i_{d}} } \sum_k
\sum_{m_{s_d}} \sum_{m'_{s_d}}\sum_{S}(2S+1)
\left(\begin{array}{ccc} s_d & s_a &
 S \\ m_{s_d} & m_{s_a} & -M_S\end{array}\right)
\left(\begin{array}{ccc} s_d' & s_a' & S \\ m'_{s_d} & m'_{s_a} &
-M_S \end{array}
\right) \nonumber \\
&\times&
(-1)^{m+m'-\omega-\omega'+M_L+m'_n-\lambda'+j+j'+n+n'+s_d+s_d'}
\sqrt{[j ][j' ][L ][L' ]}\sum_{n n'} [n ][n' ] \nonumber \\
&\times& \left(\begin{array}{ccc} j & s_d &
 n \\ m & -m_{s_d} & -m_n \end{array}\right)
\left(\begin{array}{ccc} n & k & n' \\ m_n & \nu & -m_n'
\end{array} \right)\left(\begin{array}{ccc} L & k &
 L' \\ -M_L & \nu & M_L' \end{array}\right)
\left(\begin{array}{ccc} j' & s_d' & n' \\ m' & -m'_{s_d} & -m_n'
\end{array} \right) \nonumber \\ &\times&
\left(\begin{array}{ccc} j & s_d &
 n \\ \omega & -\sigma & - \lambda \end{array}\right)
\left(\begin{array}{ccc} n & k & n' \\ \lambda & q & -\lambda' \end{array}
\right)\left(\begin{array}{ccc} L & k &
 L' \\ 0 & 0 & 0 \end{array}\right)
\left(\begin{array}{ccc} j' & s_d' & n' \\ \omega' & -\sigma' &
-\lambda' \end{array} \right)\sqrt{\frac{(2k+1)}{2}} \kappa
V_{k |q|}^{\alpha,2S+1}(R).
\end{eqnarray}
\end{widetext}
The potential matrix elements in basis sets B2p and B3p are
trivially related to the ones in B2w and B3w, respectively,
requiring only the change to a parity-symmetrized basis set, built
as a superposition of $+\lambda$ and $-\lambda$ or $+\omega$ and
$-\omega$ vectors. Finally, the evaluation of the potential in
basis set B1 can be easily obtained from that in B3p by taking the
standard composition of $j$ and $s_a$ with the respective nuclear
angular momenta $i_d$ and $i_a$.

\begin{figure}
\setlength{\unitlength}{4mm}
\centerline{\includegraphics[width=.9\linewidth,height=0.75\linewidth,angle=0]{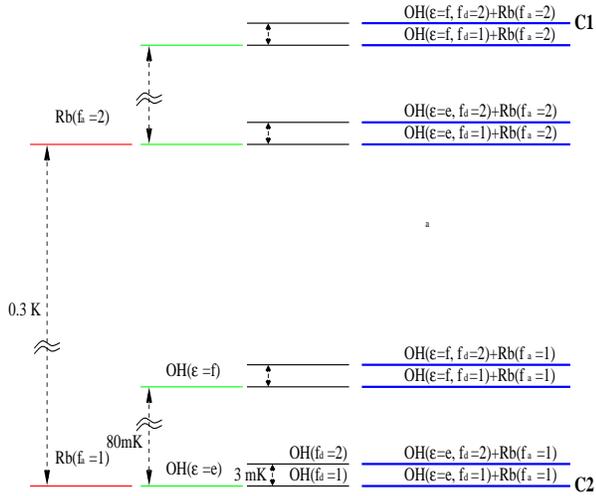}}
\caption{Threshold diagram for the channels of interest, labeled
by quantum numbers that characterize the internal states of both
colliding partners. These correspond to the ground rotational
state $\bar{\omega}=3/2, j=3/2$ of OH. Thresholds corresponding to
the incident channels considered in section \ref{CR} (C1 and C2)
are indicated. Hyperfine splitting in Rb and $\Lambda$-doubling
and hyperfine splitting in OH are given. \label{levelsh}}
\end{figure}

The true eigenstates of OH are linear combinations of functions
with different values of $\omega$, mixed by spin-uncoupling terms
in the hamiltonian. The mixing is significant even for the
rotational ground state: $85\% $ of $\omega=3/2$ and $15\% $ of
$\omega=1/2$. We have approximated OH as a pure case (a) molecule
for convenience in the present work. The fine and hyperfine
energies,  taken
from the work of Coxon {\em et al.} \cite{Coxon}, were associated to a unique
set of case (a) quantum numbers $|s_d \bar{\lambda} \bar{\omega}
\epsilon (j i_d )f_d m_{f_d} \rangle $.
For Rb, the experimental values are taken from ref.\
\onlinecite{Arim}.
 Figure \ref{levelsh}
shows the quantum numbers that characterize the internal states of
the colliding partners for the 8 lowest asymptotic thresholds,
corresponding to the rotational state $\bar{\omega}=3/2, j=3/2$ of
OH.

\vspace{-.2cm}
\subsection{Solving the Schr{\"o}dinger equation}\label{Solve}
The full Hamiltonian operator can be written
\begin{eqnarray}
\hat{H}&=&-{\hbar^2\over{2\mu}}
     \left[ R^{-1}\left(\frac{\partial^2}{\partial
R^2}\right) R - \frac{{\hat{L}}^2}{R^2} \right] + \hat{V}(R) +
\hat{H}_{\rm th},
\end{eqnarray}
where $\mu$ is the reduced mass, $\hat{V}(R)$ indicates the {\em
potential} matrix containing the electronic potential matrix
elements given in the previous section and $\hat{H}_{\rm th}$ is
the constant {\em thresholds} matrix. By
construction, any constant difference of energy between channels
has been relegated to the threshold matrix so that the potential
matrix elements die off at long range.

The coupled-channel equations that result from introducing this
Hamiltonian into the total Schr{\"o}dinger equation are propagated
using the log-derivative method of Johnson \cite{Johnson},
modified to take variable step sizes keyed to the local de Broglie
wavelength. The log-derivative matrix $Y$ thus obtained is matched
to spherical Bessel functions in the usual way to yield the
scattering matrix $S$. Using $T$ matrix elements ($T=i(S-1)$)
\cite{Mott}, the cross section for incident energy $E$ between
internal states $\alpha$ and $\beta$, corresponding to a beam
experiment, can be obtained using the expression
\begin{eqnarray}\label{nondiag}
\sigma_{\alpha \rightarrow \beta} (E) &=& \sum_{L L' M'_L L'' }
\frac{\pi}{k^2} i^{L'' - L} \sqrt{(2L+1)(2L'' +1)} \nonumber \\
&\times&T^{*} _{\alpha L0 \rightarrow \beta L' M'_L}(E) T_{\alpha
L'' 0 \rightarrow \beta L' M'_L}(E),
\end{eqnarray}
where the collision axis is chosen along the quantization axis (so
that only $M_L=0$ contributes). The labels $\alpha$ and $\beta$
designate sets of quantum numbers specifying the internal states
of both colliding partners ($s_d \bar{\lambda} \bar{\omega}
\epsilon (j i_d )f_d m_{f_d}, (s_a i_a )f_a m_{f_a}$), and
$k=(2\mu E/\hbar^2)^{1/2}$ is the wave number corresponding to
incident kinetic energy $E$.

Our aim is to extract information for collisional processes
involving OH$(j=3/2, ^2\Pi_{3/2})+{\rm Rb}(5s ^1,\ ^2S_{1/2})$, in
any of their hyperfine states, with translational energies in the
range $10^{-6}$ to $1$ K. This range comfortably includes both the
temperatures currently reached in buffer-gas or Stark deceleration
experiments and the target temperatures of sympathetic cooling. A
fully converged calculation would require propagation to large
values of $R$ and the inclusion of a huge number of channels.
Although both covalent and ion-pair channels should be considered,
in this pilot study we include only covalent states, except in
Section \ref{harp} below. The large anisotropy of the surfaces
included in the calculation makes it necessary to use a large
number of OH rotational states and partial waves; the inclusion of
rotational states up to $j=11/2$ is required (these states lie
$\approx 550$ cm$^{-1}$ ($\bar{\omega}=3/2$) and $\approx 850$
cm$^{-1}$ ($\bar{\omega}=1/2$) above the ground state, numbers
comparable to the depth of the covalent potential energy
surfaces). The number of partial waves needed for convergence for
each rotational state is shown in Table \ref{tabwaves}; it may be
seen that fewer partial waves are required for higher rotational
states. Unfortunately, a fully converged calculation was beyond
our computational resources, so we reduced the number of channels
to include only $40 \%$ of those in Table \ref{tabwaves}. This
gives cross sections accurate to within a factor of 2.

\begin{table}
\caption[table1]{ Approximate number of partial waves required in
a converged calculation for each rotational state ($\bar{\omega},
j$). }
\begin{center}
\begin{tabular}{ccccccc} \hline
&$j=1/2$&$j=3/2$&$j=5/2$&$j=7/2$ &$j=9/2$ & $j=11/2$
   \\ \hline
 $\bar{\omega}=3/2$ & & 70 & 70 & 60 & 50 & 25
\\ $\bar{\omega}=1/2$ & 70 & 70 & 70 & 65 & 50 & 40
   \\ \hline

\end{tabular}
\end{center}

\label{tabwaves}
\end{table}

We consider first the collision of atoms and molecules that are
both in their maximally stretched states, $(f_a,m_{f_a}) =
(f_d,m_{f_d})=(2,+2)$. The s-wave incident channel has $L=0$ and
$M_L=0$, so corresponds to $M_{\mathcal J}=+4$. The set of all
$M_{\mathcal J}=+4$ channels with a defined total parity $p$,
including all allowed $M_L$ projections, as well as all the $f_a$,
$f_d$, $m_{f_a}$ and $m_{f_d}$ quantum numbers (or equivalently
$m$, $m_{s_a}$ and $m_{i_a}$, $m_{i_d}$) contains 23433 channels.
This makes an exact calculation infeasible. We have therefore
introduced two approximations to reduce this number. First, the
projection $m_{f_d}$ for channels with $j>5/2$ is fixed to its
initial value; the suppressed projections increase the degeneracy
and might split rotational Feshbach resonances, but numerical
tests show that making this approximation does not substantially
alter the overall magnitude of the cross sections reported here.
This approximation reduces the number of channels to 10555.
Second, for propagation at large $R$ we disregard channels that
are ``locally closed'', that is, whose centrifugal barrier is higher than
the incident energy in a given amount that is modified until converence.

Even with these approximations, and the suppression of ion-pair
channels, it is impractical to perform full calculations. However,
dividing the radial solution of the Schr{\"o}dinger equation into
an inner region ($R<R_0$) and an outer region ($R>R_0$) makes it
possible to use different basis sets (``frames'') in each of them.
Frame transformations have previously been employed for the
simpler problem of alkaline earth+alkaline earth collisions
\cite{Burke} and
for electron-molecule collisions \cite{Fano}. They will be an
essential tool for introducing hyperfine structure into atom-molecule and
molecule-molecule collision problems \cite{Nueva}.
The calculation is thus divided into two different steps:

\begin{itemize}
\item
At short range, $R< R_0$, the hyperfine interaction is small
compared to the depths of the short-range potentials. We therefore
represent the Hamiltonian in basis set B3 (see Appendix 2), where
the potential is diagonal in nuclear spin projections $m_{i_{a}}$
and $m_{i_{d}}$. There are 8 such blocks (since
$(2i_d+1)(2i_a+1)=8$). We ignore elements of $ \hat{H}_{\rm th}$
that couple different pairs of $m_{i_{a}}$ and $m_{i_{d}}$. This
reduces a single $(8N)\times (8N)$ calculation to 8 $(N\times N)$
calculations.
At $R=R_0$ the complete $Y$ matrix can be rebuilt using the
partial $Y_{m_{i_{a}},m_{i_{d}}}$ matrices obtained from each
subset,
 $ |s_d \bar{\lambda} \bar{\omega} \epsilon j m \rangle |s_a m_a \rangle | L M_L\rangle $
\begin{widetext}
\begin{eqnarray}
 \lefteqn{\langle L M_L | \langle i_a m_{i_{a}} | \langle s_a m_{s_a} | \langle i_d m_{i_{d}} | \langle s_d \bar{\lambda} \bar{\omega} \epsilon j m | Y |s'_d \bar{\lambda}' \bar{\omega}' \epsilon' j' m' \rangle | i_d'
 m_{i_{d}}' \rangle |s_a' m'_{s_a} \rangle | i_a' m'_{i_a} \rangle| L' M_L'\rangle } \nonumber \hspace{1.5cm}\\ &=& \delta_{i_a i'_a} \delta_{m_{i_{a}} m'_{i_{a}} }
\delta_{i_d i'_d} \delta_{m_{i_{d}} m'_{i_{d}} } \langle L M_L |
\langle s_a m_{s_a} | \langle s_d \bar{\lambda} \bar{\omega}
\epsilon j m | Y_{m_{i_{a}},m_{i_{d}}} |s'_d \bar{\lambda}'
\bar{\omega}' \epsilon' j' m' \rangle |s_a' m'_{s_a} \rangle | L'
M_L'\rangle.
\end{eqnarray}
 \end{widetext}
This $Y$ matrix is then transformed into the asymptotic basis set
B1.
We have found that this frame transform provides a very accurate way
to include the Rb-OH hyperfine structure in reduced calculations
using only covalent channels. Moreover, owing to the depth of the
short-range potentials, $Y$ is weakly dependent on energy and can
be interpolated in the inner region.

\item
At long range, $R > R_0$, the $Y$ matrix, expressed already in B1,
 is propagated  to large distances to obtain the $S$ matrix. 
We invoke an alternative approximation:
The coupling between different asymptotic rotational and
fine-structure states diminishes at longer distances and can be
neglected. Thus the subset corresponding to the ground rotational
diatomic state ($j=3/2$, $\bar{\omega}=3/2$) can be propagated by
itself to asymptotic distances. We reinstate the full
hyperfine Hamiltonian in this region.
\end{itemize}

It is worth noting that basis set B2, defined in Appendix 2, could also be
the basis for another frame transformation.
 Approximate decoupling
of singlet and triplet channels in the inner region would allow a 
partition of the
numerical effort into two smaller groups of channels, and 
the introduction of the ionic channels in the singlet one.

\vspace{-.2cm}
\section{Scattering cross sections}\label{CR}
We have calculated elastic and state-resolved inelastic cross
sections for two different incident channels for collisions of Rb
atoms with OH molecules. These are shown as C1 and C2 in Figure
\ref{levelsh}. Although we do not include the effects of external
fields explicitly in this work, we consider states in which both
partners can be in weak-field-seeking states in a magnetic field,
and thus magnetically trappable. The OH hyperfine states that can
be trapped at laboratory magnetic fields are $(f_d=2, m_{f_d}=+2,+1,0)$
and $(f_d=1,m_{f_d}=+1)$, while the corresponding states for Rb
are $(f_a=2, m_{f_a}=+2,+1,0)$; the Rb state $(f_a=1, m_{f_a}=-1)$ is
 trappable for fields smaller that $\approx 1250$ gauss. 

In the first case, designated C1 in Figure \ref{levelsh}, both
partners are in maximally stretched states: OH$(\epsilon=f, f_d=2,
m_{f_d}=2)$ + Rb$(f_a=2, m_{f_a}=2)$. This case corresponds to the
highest threshold correlating with OH in its ground rotational
state. In this case, OH is also electrostatically trappable. The
second case, designated C2 in Figure \ref{levelsh}, correlates
with the lowest asymptotic threshold in the absence of an external
field: OH$(\epsilon=e, f_d=1, m_{f_d}=1)$ + Rb$(f_a=1,
m_{f_a}=-1)$. Both partners are again magnetically trappable,
although in a field they will no longer be the lowest energy
states.

\begin{figure}
\setlength{\unitlength}{4mm}
\centerline{\includegraphics[width=1.05\linewidth,height=1.2\linewidth,angle=-0]{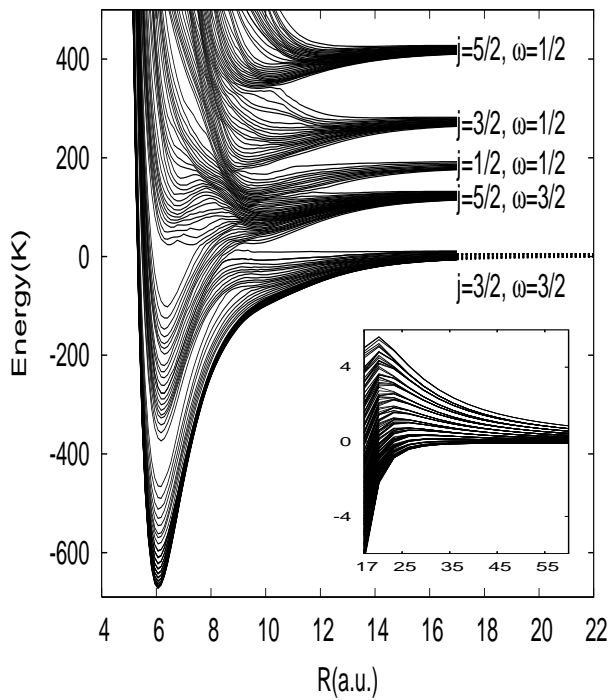}}
\hspace{-1.5cm} \caption{Adiabatic curves correlating with the
lower rotational states for the collision in maximally stretched
states of both partners ($p=+1,M_{\mathcal J}=+4$): for $R < 17\
a_0$ hyperfine interaction is partially neglected; for long range
$R > 17\ a_0$, only channels with OH in its ground rotational
state are considered and full hyperfine is included (inset gives a more
detailed image of the latter regime).
\label{adiab}}
\end{figure}

Figure \ref{adiab} shows selected adiabatic curves correlating
with the lower rotational states for the collision with both
partners in maximally stretched states ($p=+1,M_{\mathcal J}=+4$).
In our calculations we take $R_0=17\ a_0$. As described above, for
$R < R_0$ the hyperfine interaction is partially neglected. For $R
> R_0$, only hyperfine channels with OH in its ground rotational
state are included.

\vspace{-.2cm}
\subsection{General behavior: total cross sections}

\begin{figure}[ht]
\setlength{\unitlength}{8mm}

\begin{picture}(-8,15.0)(14,0)

\put( 5, 9){\
\includegraphics[width=.85\linewidth,height=0.95\linewidth,angle=-90]{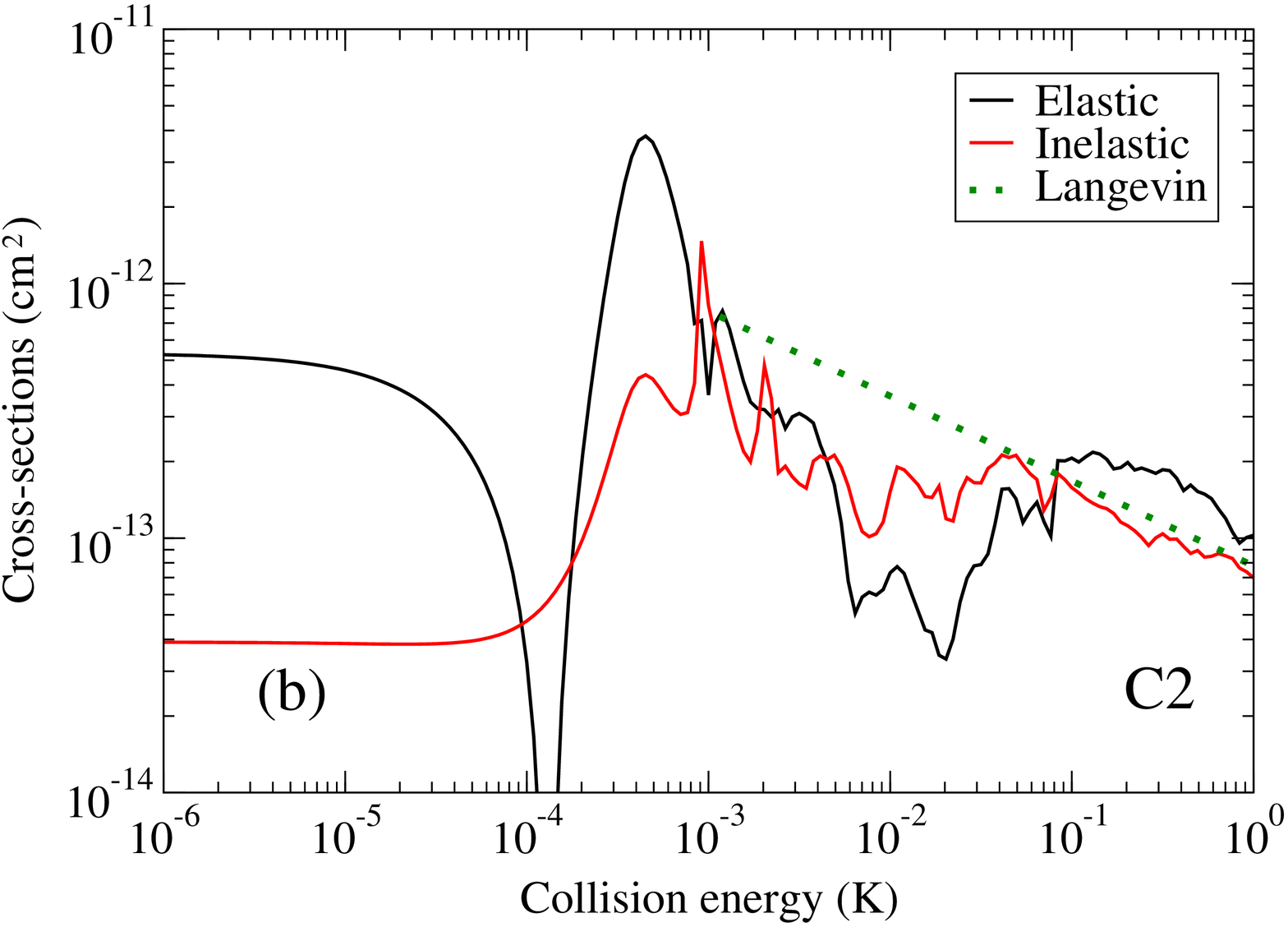}}
\put( 5, 17){\
\includegraphics[width=.85\linewidth,height=0.95\linewidth,angle=-90]{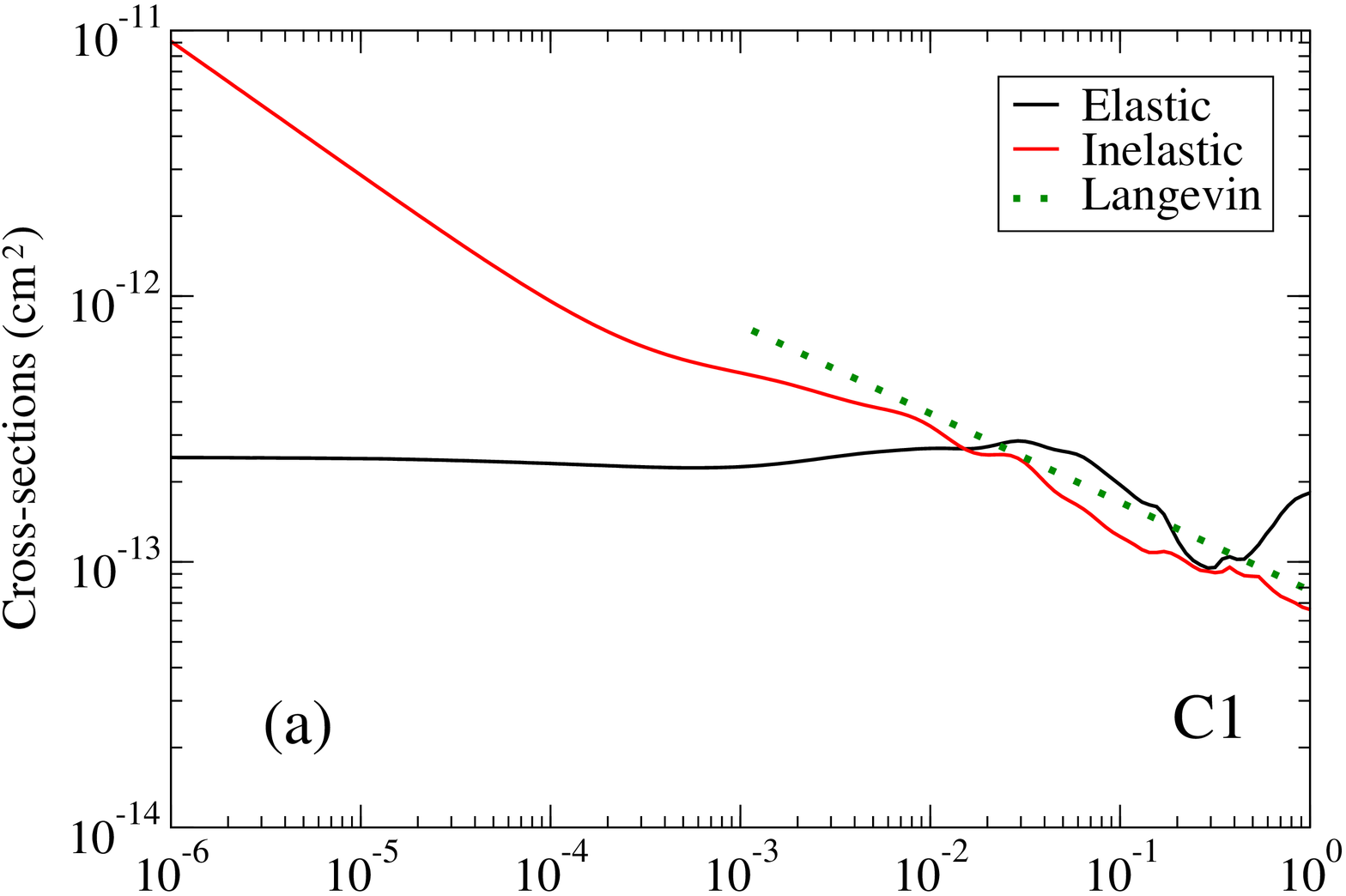}}

\end{picture}
\caption{Elastic and total inelastic  cross sections for the cases
analyzed: Panel (a) corresponds to OH$(\epsilon=f, f_d=2,
m_{f_d}=2)$ + Rb$(f_a=2, m_{f_a}=2)$ incident channel (C1). Panel
(b) corresponds to OH$(\epsilon=e, f_d=1, m_{f_d}=1)$ + Rb$(f_a=1,
m_{f_a}=-1)$ incident channel (C2). The points indicate the
Langevin cross section. \label{crosses0}}
\end{figure}

We begin by showing in Figure \ref{crosses0} the total cross
sections for incident channels C1 and C2. A brief description of
these has been reported previously \cite{PRL}. Below $10^{-4}$~K
for incident channel C1, or $10^{-5}$~K for C2, the Wigner
threshold law applies. Namely, as the energy goes to zero, cross
sections corresponding to elastic and isoenergetic processes
approach a constant value, while those for exoergic processes vary
as $1/\sqrt{E}$, rapidly exceeding elastic cross sections. No
quantitative predictive power is expected in this region. Rather,
the values of threshold cross sections are strongly subject to
details of the PES, and are typically only
uncovered by experiments.

At higher energies, above $10^{-2}$\ K, where many partial waves
contribute ($l \ge 4$ for C1 and $l \ge 5$ for C2), the behavior
of the cross sections changes and inelastic processes are well
described by a semiclassical Langevin capture model \cite{17,
PRL},
\begin{eqnarray}\sigma_{\rm Langevin}(E)=3 \pi \left( \frac{C_6}{4E}
\right)^{1/3}.
\end{eqnarray}
This cross section is also plotted in Figure \ref{crosses0}
(points). The Langevin expression arises as a limit of the exact
quantum expression in Eq.\ \ref{nondiag}, with the usual
assumptions: the impact parameter takes values in a continuous
range, and the height of the centrifugal barrier, determined using
only long-range behavior, determines the number of partial waves
that contribute for a given energy. Similar behavior has been
observed previously in cold collisions, such as $M + M_2$ with $M$
an alkali metal. For Li + Li$_2$, $l>3$ was found to be a
sufficient for the cross sections to exhibit Langevin behavior
\cite{Li2homo}.

As can be seen in Figure \ref{crosses0}, the Langevin limit
reproduces the general trend across the entire semiclassical
energy range. In a Hund's case (a) system like OH, where the
electron spin is strongly tied to the intermolecular axis, the
highly anisotropic PES might be expected to
disrupt the spin orientation relative to the laboratory frame
completely. As a consequence, inelastic processes are expected to
be very likely and the Langevin model should describe well the
behavior of Rb-OH and similar systems. A similar upper limit for
the elastic cross section is given by four times the inelastic
one.
It is easy to verify that, if the inelastic cross section reaches
its maximum value, the elastic and inelastic contributions to the
cross section must be equal. This behavior is also seen in Figure
\ref{crosses0}.

The cross sections for incident channel C2 are quite different
from those for incident channel C1. For C2, the cross sections are
highly structured, exhibiting a large number of Feshbach
resonances. Since the atom and the diatom are both in their
lowest-energy state, there are plenty of higher-lying hyperfine
states to resonate with. However, the pronounced minimum in the
elastic cross section at $E \sim 10^{-4}$~K is the consequence of
a near-zero s-wave phase shift.

\vspace{1.2cm}

\vspace{-.2cm}
\subsection{Detailed picture: partial cross sections}

Some additional insight into the collision process can be obtained
by examining the partial cross sections to various final states.
However, there are many of these. Starting in incident channel C1,
there are 128 possible outcomes, counting the hyperfine states of
both Rb and OH, plus the lambda doublet of OH. To simplify this
information, we first break the cross sections into four classes:
elastic scattering, scattering in which only the Rb state changes,
scattering in which only the OH state changes, and scattering in
which both change.

\begin{figure}[ht]
\setlength{\unitlength}{8mm}

\begin{picture}(-8,15.0)(14,0)

\put( 5, 9){\
\includegraphics[width=.85\linewidth,height=0.95\linewidth,angle=-90]{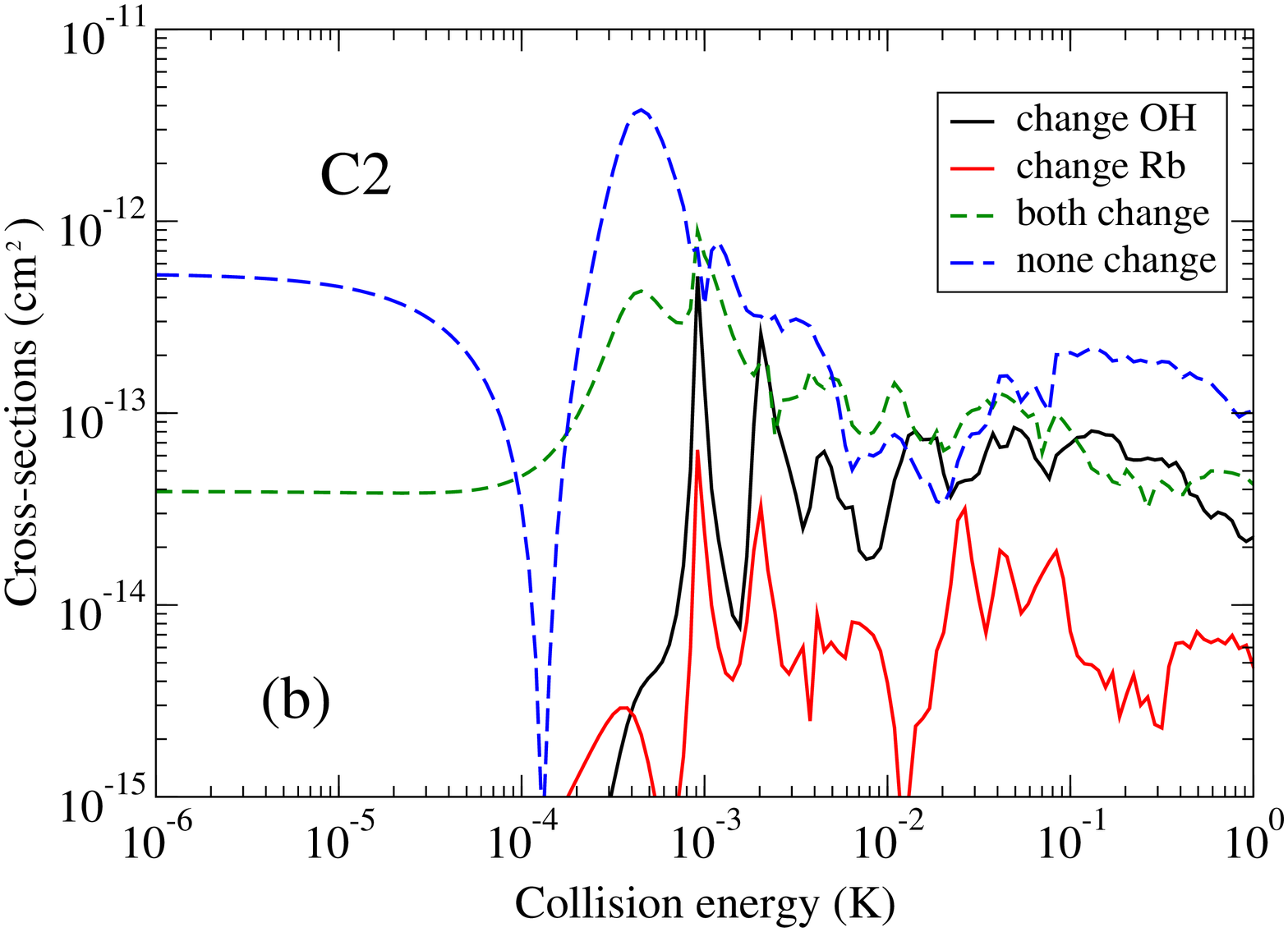}}
\put( 5, 17){\
\includegraphics[width=.85\linewidth,height=0.95\linewidth,angle=-90]{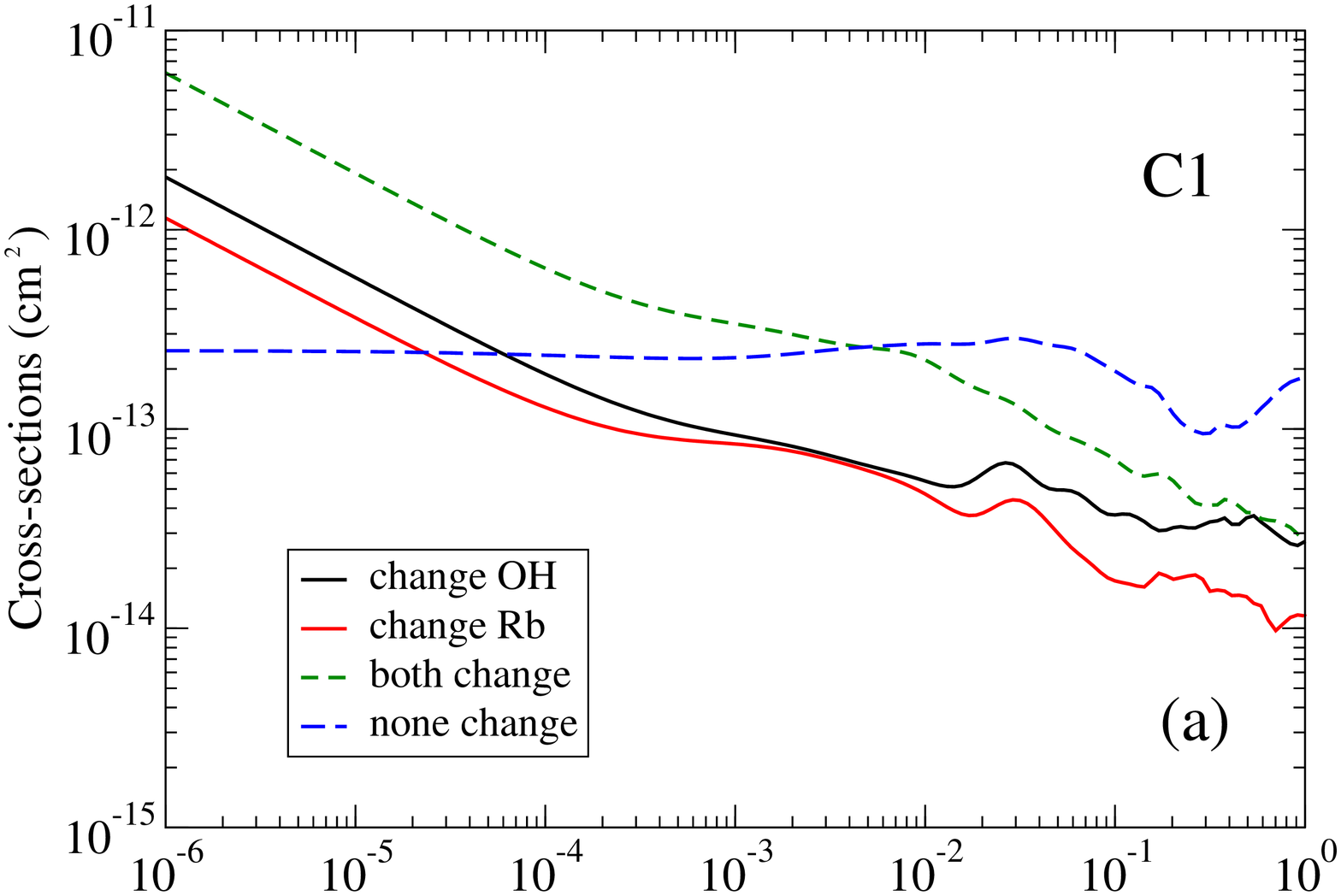}}

\end{picture}
\caption{ Total cross sections for 4 types of process are
considered: those where only OH partner changes internal state,
those where only Rb changes, those where both partners suffer
inelastic change and those where neither change. Panel (a)
corresponds to OH$(\epsilon=f, f_d=2, m_{f_d}=2)$ + Rb$(f_a=2,
m_{f_a}=2)$ incident channel (C1). Panel (b) corresponds to
OH$(\epsilon=e, f_d=1, m_{f_d}=1)$ + Rb$(f_a=1, m_{f_a}=-1)$
incident channel (C2). \label{crosses2}}

\end{figure}

These four possibilities are shown in Figure \ref{crosses2} for
both incident channels, C1 and C2. In general, all four processes
are likely to occur. This further attests to the complete
disruption of the spin during a collision. Nevertheless, at the
highest energies probed (where results are less sensitive to
potential details) there is a definite propensity for the OH
molecule to change its internal state more readily than the Rb
atom, at least when only one of them changes. This is probably a
consequence of the spherical symmetry of the Rb atom, whereby its
electronic spin is indifferent to its orientation. By contrast,
the electronic angular momentum of OH is strongly coupled to the
molecular axis, and will follow its changes in orientation due to
the anisotropies in the interaction.

A more detailed understanding can be obtained by considering Eq.\
\ref{elements}. For Rb the hyperfine projection is given by
$m_{f_a} = m_{s_a} + m_{i_a}$, whereas for OH it is given by
$m_{f_d} = m_{n}+m_{s_d}+m_{i_d}$. The nuclear spin projections
$m_{i_a}$ and $m_{i_d}$ are untouched by potential energy
couplings. The potential conserves $M_S=m_{s_a}+m_{s_d}$, so that
if the Rb electronic spin $m_{s_a}$ changes, so will the OH
electronic spin $m_{s_d}$. On the other hand, OH can also change
the projection $m_n$ of the rotational angular momentum $n$, which
is absent in Rb. Thus OH has more opportunities to change its
internal state than does Rb, and this is reflected in the
propensities in Figure \ref{crosses2}.

\begin{figure}[ht]
\setlength{\unitlength}{4mm}
\centerline{\includegraphics[width=.85\linewidth,height=0.95\linewidth,angle=-90]{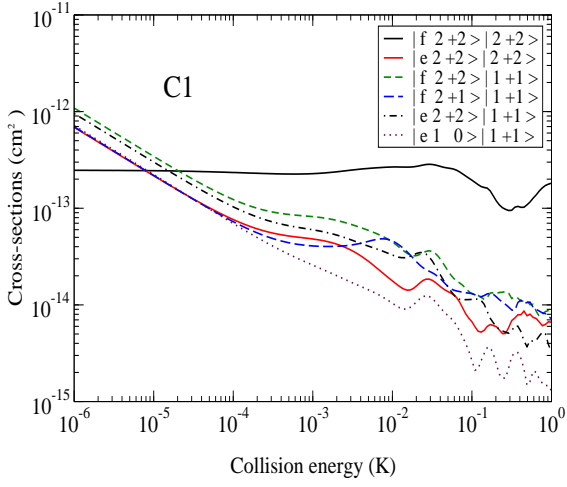}}
\caption{Dominant partial cross sections for OH$(\epsilon=f,
f_d=2, m_{f_d}=2)$ + Rb$(f_a=2, m_{f_a}=2)$ incident channel (C1).
\label{crosses}}

\end{figure}

We first consider incident channel C1, with both collision
partners initially in their maximally stretched states. Some of
the main outcomes are shown in Fig \ref{crosses}. In the absence
of an anisotropic interaction, and remembering that
$M_{\mathcal{J}}=m_{f_a}+m_{f_d}+M_L$ is conserved, the sum of the
spin projections $M_F=m_{f_a}+m_{f_d}$ would be conserved. Thus
inelastic collisions would be impossible. This is because the
projection of $m_{f_d}$ could not be lowered without raising
$m_{f_a}$, but $m_{f_a}$ could not be raised further. However, the
anisotropic PES in Rb-OH allows such changes
quite readily.

\begin{figure}[ht]
\setlength{\unitlength}{8mm}

\begin{picture}(-8,17.0)(14,0)

\put( 5, 9){\
\includegraphics[width=.85\linewidth,height=0.95\linewidth,angle=-90]{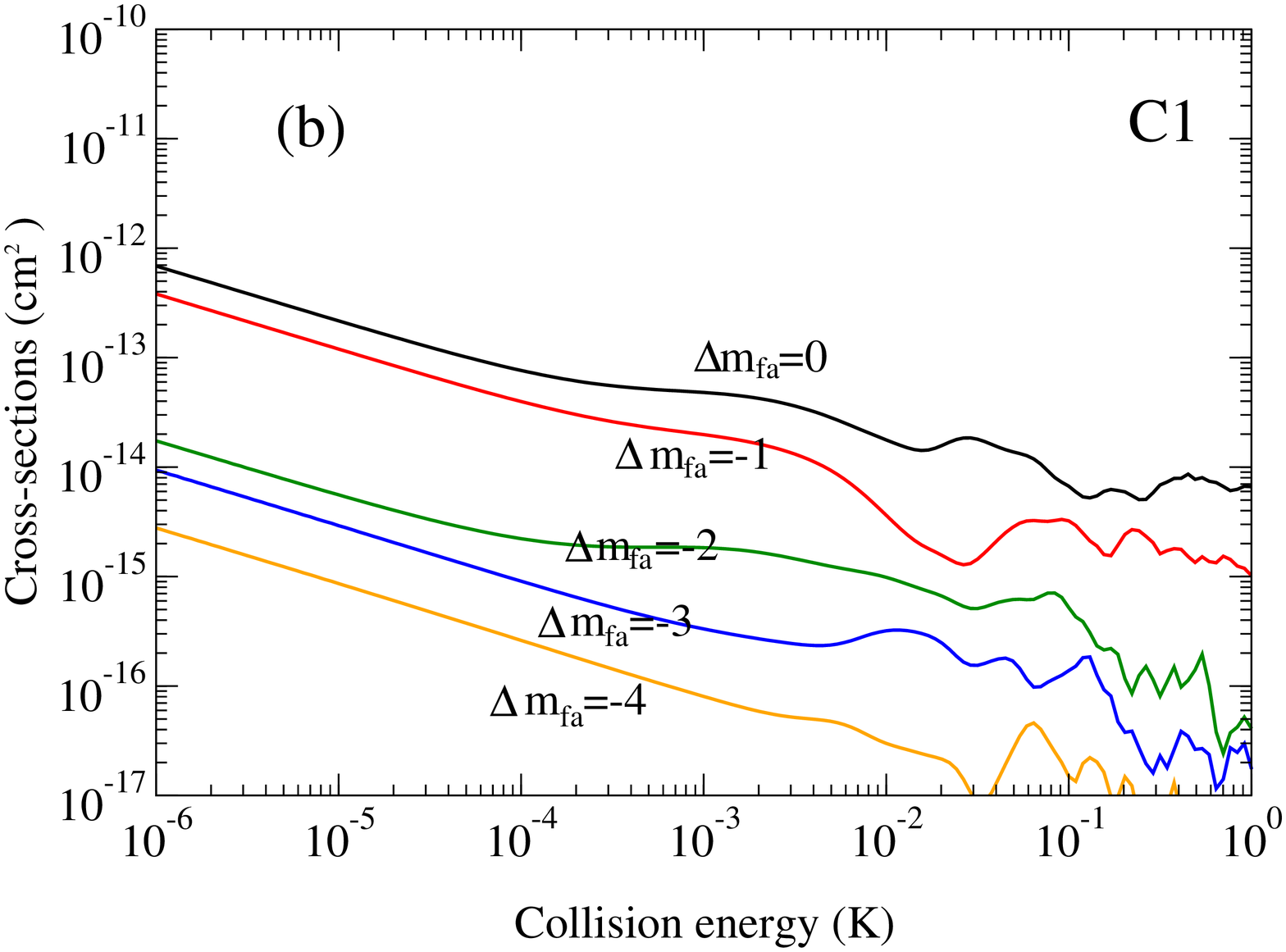}}
\put( 5, 17){\
\includegraphics[width=.85\linewidth,height=0.95\linewidth,angle=-90]{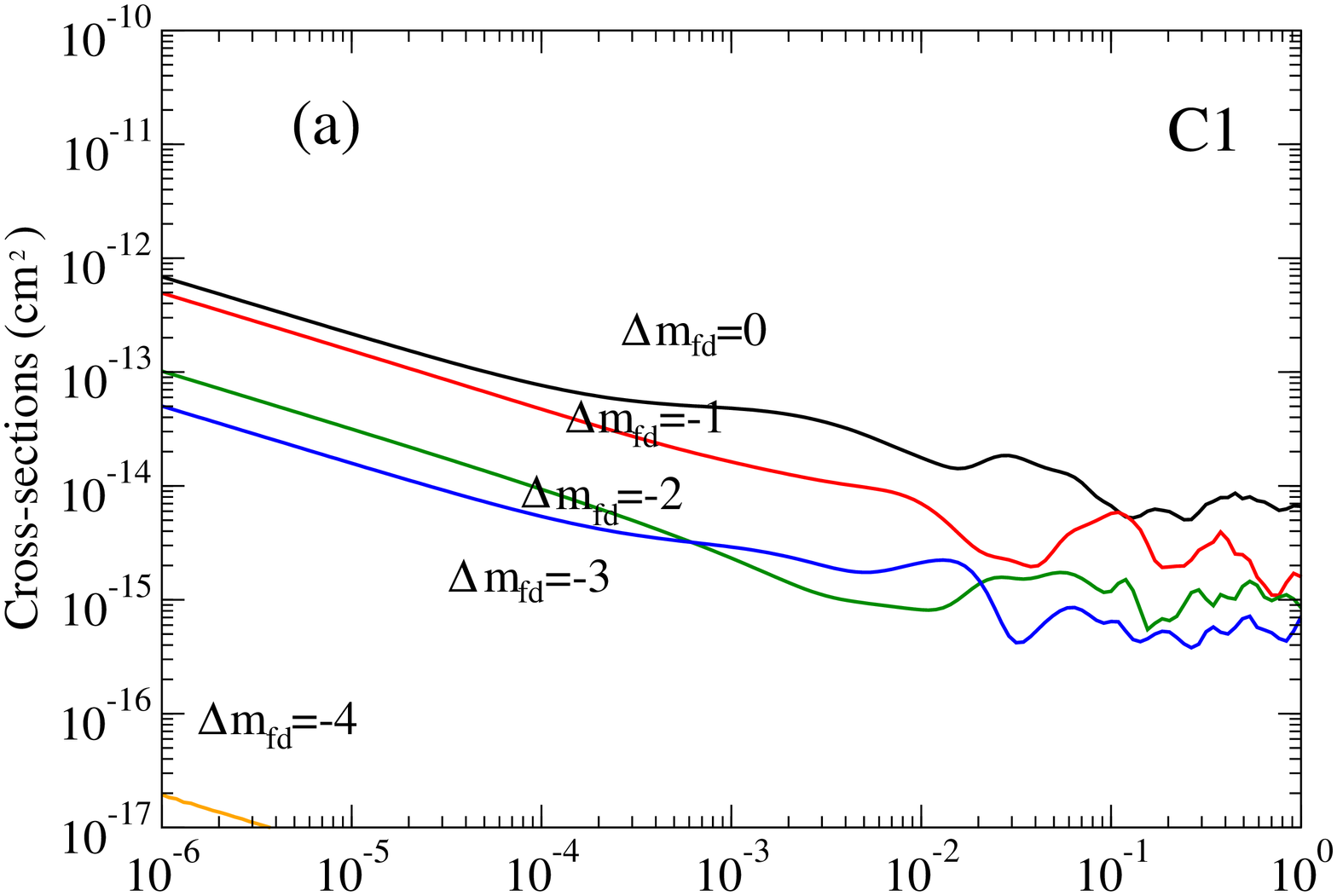}}

\end{picture}
\caption{Panel (a) ((b)): inelastic cross sections for $m_{f_d}$
($m_{f_a}$) changing collisions. The higher the change, the lower
the cross section. They correspond to OH$(\epsilon=f, f_d=2,
m_{f_d}=2)$ + Rb$(f_a=2, m_{f_a}=2)$ incident channel (C1).
\label{OHbaja}}
\end{figure}

There remains, however, a propensity for collisions with small
values of $\Delta m_f$ to be more likely, as seen in Figure
\ref{OHbaja}. Part (a) shows cross sections that change $m_{f_d}$
for OH without changing $m_{f_a}$ for Rb, while part (b) shows
those that change $m_{f_a}$ without changing $m_{f_d}$. As noted
above, Rb appears to be more reluctant to change its projection.
In fact, since the potential is diagonal in the states of the
nuclei, only consecutive values of $m_{f_a}$ are coupled in first
order. On the other hand, first-order coupling exists between many
different values of $m_{f_d}$.
Finally, a decrease in the probability of processes with
increasing $\Delta M_F$ can be related to the diminution of
anisotropic terms in the potential when increasing the angular
momentum transfer ($k$ in Eq.\ \ref{elements}).

\begin{figure}[ht]
\setlength{\unitlength}{8mm}

\begin{picture}(-8,17.0)(14,0)

\put( 5, 9){\
\includegraphics[width=.85\linewidth,height=0.95\linewidth,angle=-90]{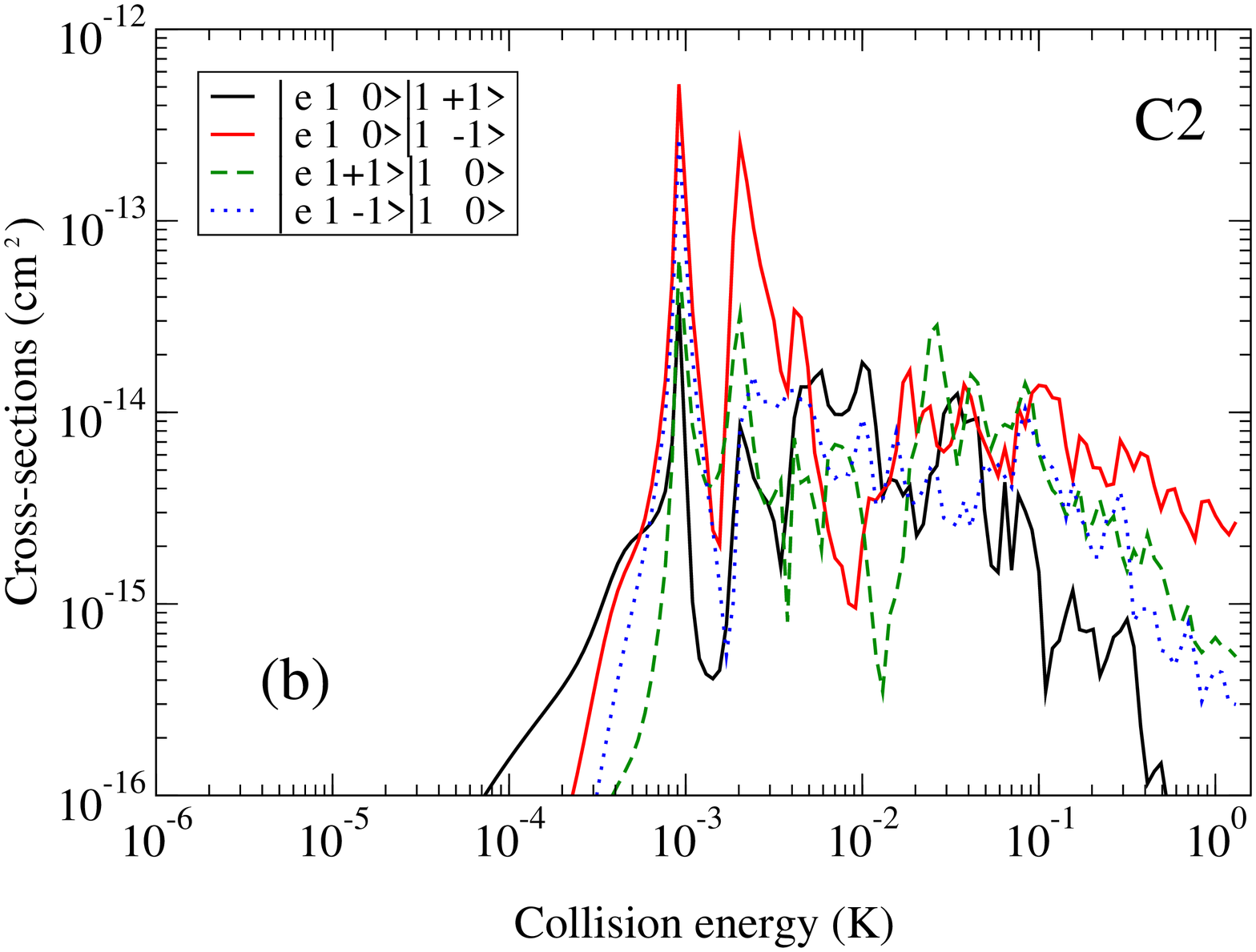}}
\put( 5, 17){\
\includegraphics[width=.85\linewidth,height=0.95\linewidth,angle=-90]{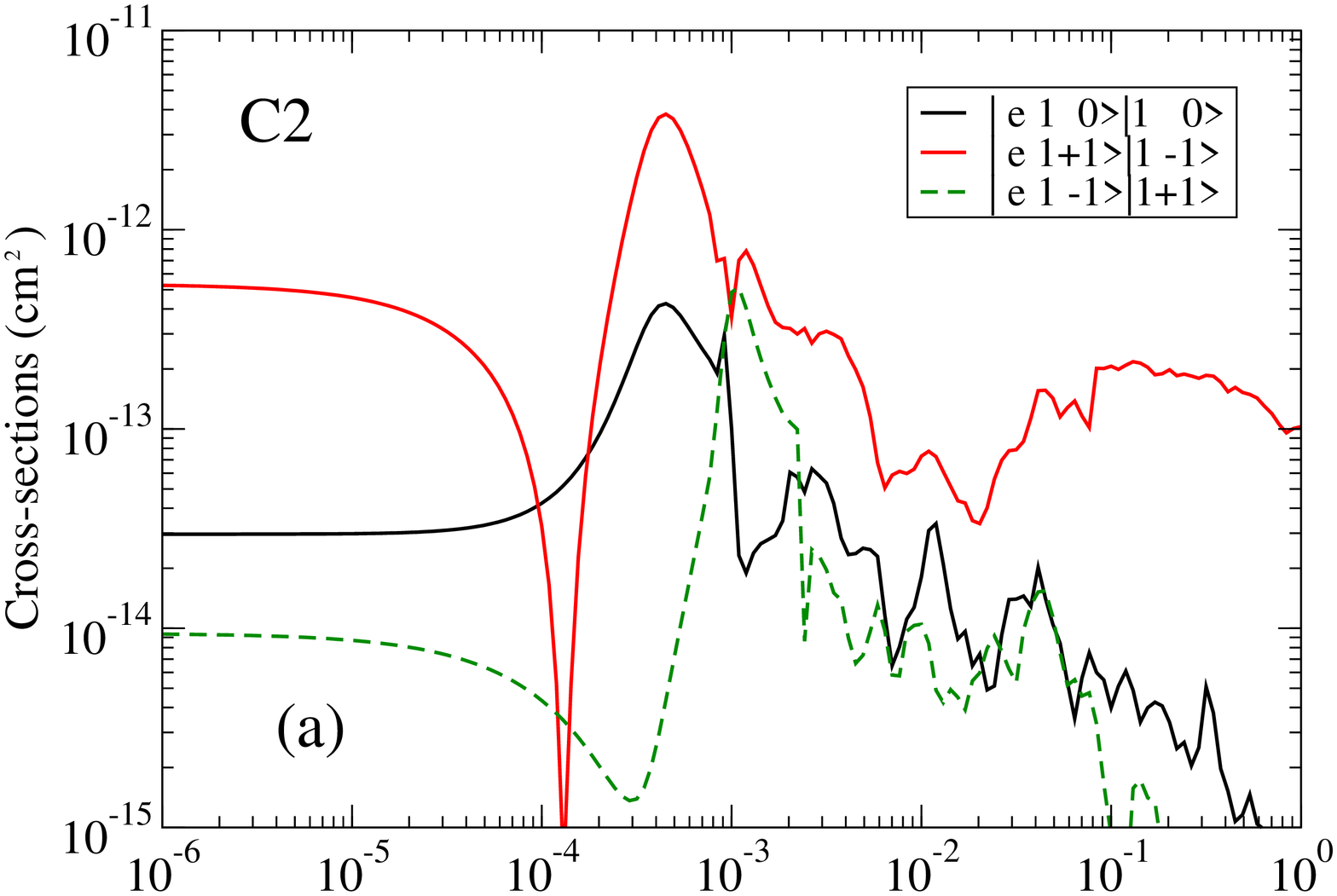}}

\end{picture}

\caption{Some partial cross sections for C2 incident channel:
OH$(\epsilon=e, f_d=1,
m_{f_d}=1)$ + Rb$(f_a=1, m_{f_a}=-1)$. On panel (a) elastic cross
section together with partial cross sections for two degenerate
processes with the initial channel, opened at zero collision energy,
are shown. The processes shown on panel (b) are also degenerate but
closed by the centrifugal barrier at zero collision
energy.\label{crosses3}}

\end{figure}

We now consider incident channel C2, OH$(\epsilon=e, f_d=1,
m_{f_d}=1)$ + Rb$(f_a=1, m_{f_a}=-1)$. This is the lowest
threshold in the absence of a field. Only three channels,
degenerate with the initial channel, are possible outcomes at very
low energies: $ |\epsilon=e, 1,{+1} \rangle |1,{-1} \rangle $ $
|\epsilon=e, 1,0 \rangle |1,0 \rangle $, and $|\epsilon=e, 1,{-1}
\rangle |1,{+1} \rangle $. Since these states have the same value
of $M_F=m_{f_d}+m_{f_a}$ as the initial channel, the processes can
occur by ordinary spin exchange with no centrifugal barrier.
Partial cross sections for these three channels are shown in
Figure \ref{crosses3} (a) over the entire energy range.
Intriguingly, inelastic (state-changing) collisions seem to be
somewhat suppressed relative to elastic scattering. Suppressed
spin-exchange rates would presumably require delicate cancellation
between singlet and triplet phase shifts \cite{Burke98}. That such
a cancellation occurs in a highly multichannel process is somewhat
unexpected. However, examining the matrix elements of the
potential reveals that there is no direct coupling between the
initial state $ |\epsilon=e, 1,{+1} \rangle |1,{-1} \rangle $ and
final state $ |\epsilon=e, 1,{-1} \rangle |1,{+1} \rangle $.
Transitions via potential couplings are therefore a second-order
process requiring the mediation of other channels.

Many other exit channels are possible, once energy and angular
momentum considerations permit them. Cross sections for several
such processes are shown in Figure \ref{crosses3} (b). For
example, the channels OH$(\epsilon=e, 1, 0)$ + Rb$(1, +1)$ and
OH$(\epsilon=e, 1, 0)$ + Rb$(1, -1)$ are not connected to the
initial channel by spin exchange, since $M_F$ changes by $-1$. In
this case, angular momentum shunts from the molecule into the
partial-wave degree of freedom, necessitating an $L=1$ partial
wave in the exit channel. Therefore, this process is suppressed
for energies below the $p$-wave centrifugal barrier, whose height
is 1.6 mK.

\begin{figure}[ht]
\setlength{\unitlength}{4mm}
\centerline{\includegraphics[width=.85\linewidth,height=0.95\linewidth,angle=-90]{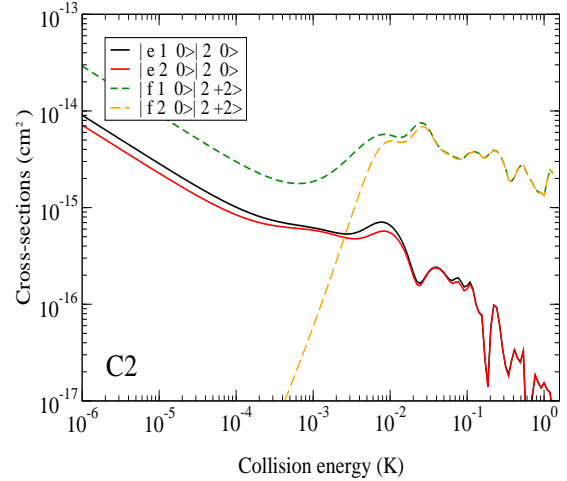}}
\caption{Partial cross sections for two pairs of processes
corresponding to incident channel C1 (OH$(\epsilon=f, f_d=2,
m_{f_d}=2)$ + Rb$(f_a=2, m_{f_a}=2)$) and final channels differing
only in the $f_d$ quantum number. Their values converge at high
kinetic energies but are very different in the cold regime.
\label{pairs}}

\end{figure}

We conclude this subsection by stressing the vital importance of
including the hyperfine structure in these calculations. Figure
\ref{pairs} shows partial cross sections for incident channel C1
scattering into pairs of channels which differ only in the small hyperfine
splitting of OH ($\approx 3$~mK). In one example (black and red
solid curves), the parity of OH changes from $f$ to $e$, leading to the
final channels OH$(\epsilon=e, f_d=1,2 , m_{f_d}=0)$ + Rb$(f_a=2,
m_{f_a}=0)$. These channels, distinguished only by their hyperfine
quantum number $f_d$, are almost identical at high energies, but
quite different at low energies. As a second example, consider the
final channels OH$(\epsilon=f, f_d=1,2 , m_{f_d}=0)$ + Rb$(f_a=2,
m_{f_a}=+2)$, (green and orange dashed), which preserve the initial parity.
The process with $f_d=1$ is exothermic, while the one with $f_d=2$
requires the opening of the partial wave threshold. Only after
both channels are open, at higher energies, do the cross sections
become almost identical.

\subsection{The harpooning process}\label{harp}

We have not yet fully incorporated the ion-pair channel in our
calculations, but it is instructive to assess its influence. In
the harpooning model it is usually understood that, if the
electron transfer takes place at long range, large collision cross
sections result \cite{Ber}. However, for Rb-OH the crossing point
$R_0 \approx 6$~\AA\ gives a geometric cross section of only
$\sigma=4 \pi R_0^2 \approx 4.5 \cdot 10^{-14}$ cm$^2$. This is
substantially smaller than the quantum-mechanical cross sections
we have estimated. Thus while harpooning may enhance cross
sections substantially at thermal energies, it is likely to have
less overall influence in cold collisions.

The inelastic results above place cross sections near the
semiclassical Langevin upper limit. The harpooning mechanism is
unlikely to increase their magnitude, but could change their
details. For example, it might be expected that the harpooning
mechanism would distribute the total probability more evenly among
the different spin orientations.

\begin{figure}[ht]
\setlength{\unitlength}{4mm}
\centerline{\includegraphics[width=.85\linewidth,height=0.95\linewidth,angle=-90]{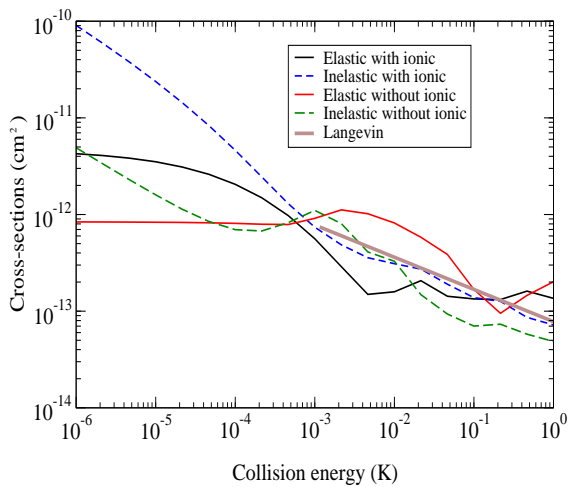}}
\caption{Effect of the inclusion of ion-pair channels in a reduced
(neglecting hyperfine structure) calculation. Langevin cross
section is also shown for comparison. \label{ionico}}

\end{figure}

We have explored the influence of the harpooning mechanism in a
reduced calculation where the hyperfine structure is neglected.
This makes the calculation tractable even with the ion-pair
channel included.
The resulting elastic and total inelastic cross sections, for incident channel
OH$(\epsilon=f, j=3/2,
m=3/2)$ + Rb$(s_a=1/2, m_{s_a}=1/2)$ (maximally stretched in the absence of
nuclear degrees of freedom)  are shown in
Figure \ref{ionico}. They are compared to the results obtained
without the ion-pair channels. In the semiclassical region, $E
> 1$~mK, the general order of magnitude of the cross sections is
preserved, although the detailed features are not. In the Wigner
regime, by contrast, including the ion-pair channel makes a quite
significant change, because completely different phase shifts are
generated in low-lying partial waves. We stress again, however,
that we do not expect this or any model to have predictive power
for the values of low-energy cross sections until the ambiguities
in absolute scattering lengths are resolved by experiments.

\section{Conclusions and prospects}

Based on the results above, we can draw several general
conclusions concerning both the feasibility of observing Rb-OH
collisions experimentally in a beam experiment and the possibility
of sympathetic cooling of OH using ultracold Rb.

We can assert with some confidence that Rb-OH cross sections at
energies of tens of mK, typical of Stark decelerators, will be on
the order of $10^{-13}\ \rm{cm}^{2}$. This is probably large
enough to make the collisions observable. In round numbers,
consider a Rb MOT with density $n_{\rm Rb} = 10^{10}$ cm$^{-3}$,
and an OH packet emerging from a Stark decelerator at 10 m/s, with
density $n_{\rm OH} = 10^{7}$ cm$^{-3}$. Further assume that the
Rb is stored in an elongated MOT that allows an interaction
distance of 1 cm as the OH passes through. In this case, a cross
section $\sigma = 10^{-13}$ cm$^2$ implies that approximately
$10^{-3}$ of the OH molecules are scattered out of the packet.
This quantity, while small, should be observable in repeated shots
of the experiment \cite{Yun}.

As for sympathetic cooling OH using cold Rb, it seems unlikely
that this will be possible for species in their stretched states
as in channel C1. The inelastic rates are almost always comparable
to, or greater than, the elastic rates. The situation becomes even
worse as the temperature drops and cross sections for exoergic
processes diverge. Every collision event {\it might} serve to
thermalize the OH gas, but is equally likely to remove the
molecule from the trap altogether, or contribute to heating the
gas.

For collision partners in the low-energy states of incident
channel C2, the situation is not as bleak, at least in the
low-energy limit, since inelastic cross sections do not diverge
and may be fairly small. However, we have focused on the
weak-magnetic-field seeking state of OH, which will not be the
lowest-energy state in a magnetic field. Thus exoergic processes
again appear, and inelastic rates will again become unacceptably
large. Under these circumstances, rather than ``sympathetic
cooling,'' the gas would exhibit ``simply pathetic cooling.''

From these considerations, it seems that the only way to guarantee
inelastic rates sufficiently low to afford sympathetic cooling
would be to remove all exoergic inelastic channels altogether. Thus both
species, atom and molecule, should be trapped in their absolute
ground states, using optical or microwave dipole traps
\cite{Mille} or possibly an alternating current trap
\cite{actrap}. Sympathetic cooling would then be forced due to the
absence of any possible outcome other than elastic and endoergic
processes. The latter might produce a certain population of
hyperfine-excited OH molecules, which could then give up their
internal energy again and contribute towards heating. However, the
molecules are far more likely to collide with atoms than with
other molecules, so that this energy would eventually be carried
away as the Rb is cooled.

Further aspects of the Rb-OH collision problem need further
work. Foremost among these is the possibility that inelastic rates
could be reduced or controlled by applying electric and/or
magnetic fields. This may result partly from the simple act of
moving Feshbach resonances to different energies, or from altering
the effective coupling between incident and final channels, as has
been hinted at previously \cite{Tick, Krems}. In addition, the
influence of the ion-pair channel may be significant. There is the
possibility, for example, that the highly polar RbOH molecule
might be produced by absorption followed by either spontaneous or
stimulated emission \cite{9}. Finally, new phenomena are possible,
such as analogues of the field-linked states that have been
predicted in dipole-dipole collisions \cite{31a, 31b}.

\acknowledgements
ML and JLB gratefully acknowledge support from the NSF and the W.
M. Keck Foundation, and PS from the M\v{S}MT \v{C}R (grant No.
LC06002) \label{sec4}

\subsection*{Appendix 1}

As described above, we encountered difficulties with
non-degeneracies between the $^1A'$ and $^1\App$ components of the
$^1\Pi$ state at linear geometries. These could have been avoided
by carrying out the linear calculations in $C_{2v}$ rather than
$C_s$ symmetry, but $C_s$ symmetry was essential to avoid
discontinuities at $\theta=0$ and $180^\circ$.

To understand the non-degeneracies it is essential to understand
the procedure used to generate the basis set used in the CISD
calculation.
\begin{enumerate}
\item The molecular orbital basis set is partitioned into internal
and external sets, and the internal orbitals are in turn
partitioned into closed and active sets. \item A reference space
is generated, including all possible arrangements of the $N$
available electrons among the active orbitals that give states of
the specified symmetry and spin multiplicity. The closed orbitals
are doubly occupied in all reference configurations. \item A set
of $(N$--2)-electron states is generated by all possible
2-electron annihilations from the reference states (with no
symmetry constraints). Some of the closed orbitals may be
designated as core orbitals, in which case annihilations from them
are not included. \item A set of $(N$--1)-electron states is
generated by all possible 1-electron additions to internal
orbitals of the $(N$--2)-electron states (with no symmetry
constraints). \item The final CI basis includes states of 3
different classes:
\begin{itemize} \item {\it Internal} states are generated by all
possible 1-electron additions to the $(N$--1)-electron states,
with the extra electron in an internal orbital, that give states
of the specified symmetry and spin multiplicity; \item {\it Singly
external} states are generated by all possible 1-electron
additions to the $(N$--1)-electron states, with the extra electron
in an external orbital, that give states of the specified symmetry
and spin multiplicity; \item Internally contracted {\it doubly
external} states are generated by 2-electron excitations into the
external space from a reference function obtained by solving a
small CI problem in the reference space.
\end{itemize}
\end{enumerate}

The non-degeneracies between the $^1A'$ and $^1\App$ components of
the $^1\Pi$ state arise from two different sources.

First, there is a part of the non-degeneracy due to the internal
and singly external configurations. At linear geometries there are
both $\Sigma$-type and $\Pi$-type reference configurations. Both
types are included for $A'$ symmetry, but only the $\Pi$-type
reference configurations are included for $\App$ symmetry. In the
$A'$ calculation, therefore, there are additional $\Pi$-type
internal and singly external configurations that arise from
$\Sigma$-type reference configurations. This effect could in
principle be suppressed in MOLPRO by including reference
configurations of both symmetries in both calculations at all
geometries, but this was prohibitively expensive in computer time.
In addition, as described below, it is responsible for only about
20\% of the total non-degeneracy.

Secondly, there is a part of the non-degeneracy due to the
(contracted) doubly external configurations. It must be remembered
that we wish to transform the $1A'$ and $2A'$ singlet states into
a diabatic representation. To do this meaningfully, we need them
to be calculated using identical basis sets. If we do the MRCI
calculations separately, this is not true: even if the reference
space is the same, each state produces a {\it different} set of
internally contracted doubly external states. It is thus necessary
to calculate the two $^1A'$ states in the {\it same} MRCI block,
so that the basis set contains both sets of contracted functions.
However, this results in a larger and more flexible basis set of
contracted doubly external functions than is generated in the
$^1\App$ case.

It is helpful to document the magnitude of the non-degeneracies in
various possible calculations. For RbOH at $\theta=0$ and
$R=12$\,\AA:
\begin{itemize}
\item If calculations are carried out in $C_{2v}$ symmetry, the
ion-pair and covalent states have different symmetry and the
non-degeneracy is not present; \item For $C_s$ symmetry with the
two $^1A'$ states calculated in a single MRCI block, both sources
of non-degeneracy are present and the non-degeneracy is
$22.78\,\mu E_h$; \item For $C_s$ symmetry with the two $^1A'$
states calculated in separate MRCI blocks, only the first source
of non-degeneracy is present and the non-degeneracy is $4.08\,\mu
E_h$; \item For $C_s$ symmetry with the two $^1A'$ states
calculated in a single MRCI block, but reference configurations of
both symmetries included, only the second source of non-degeneracy
is present and the non-degeneracy is $18.70\,\mu E_h$; \item For
$C_s$ symmetry with the two $^1A'$ states calculated in separate
MRCI blocks, with reference configurations of both symmetries
included, there is no non-degeneracy.
\end{itemize}

\subsection*{Appendix 2}
The interaction potential does not involve the spins of the nuclei
and is diagonal in the total electronic spin $S$. This makes it
expedient to define two additional basis sets.

Basis set B2 is based on Hund's case (b) quantum numbers for the
molecule: the orbital angular momentum of the electron and the
rotational angular momentum couple to form $n$, with $\lambda$
projection on the internuclear axis. $n$ coupled to the spin of
the electron $s_d$ would give us $j$ for the diatomic fragment, as
in basis set B1. Instead, in B2 we couple $s_d$ with $s_a$ to
obtain the total spin of the electrons, described by kets $|S M_S
\rangle$. However, we leave aside the states of the nuclei, which
are represented by $ | i_a m_{i_{a}} \rangle | i_d m_{i_{d}}
\rangle$, with $m_{i_{a}}$ and $ m_{i_{d}} $ being the projections
of the nuclear spin of the atom and diatom in the laboratory
frame. Basis set B2w is then given by
\begin{eqnarray} | B2w \rangle = | (s_d s_a) S M_S \rangle | n m_n \lambda \rangle | i_d
 m_{i_{d}} \rangle | i_a m_{i_{a}} \rangle
| L M_L \rangle,
\end{eqnarray}
where $ |n m_n \lambda \rangle$ represents $\sqrt{\frac{2n+1}{8
\pi^2}} D_{m_n \lambda }^{n*}(\alpha,\beta,\gamma)$. The parity
operator $E^{*}$ acts on these states as follows:
 \begin{eqnarray} E^{*} | (s_d s_a) S M_S \rangle | n m_n \lambda \rangle | i_d
 m_{i_{d}} \rangle | i_a m_{i_{a}} \rangle
| L M_L \rangle=  \nonumber \\ (-1)^{n+L} | (s_d s_a) S M_S
\rangle | n m_n -\lambda \rangle | i_d
 m_{i_{d}} \rangle | i_a m_{i_{a}} \rangle
| L M_L \rangle,\nonumber \\
\end{eqnarray}
which allows us to construct combinations of define parity,
 \begin{eqnarray} | B2p \rangle = | (s_d s_a) S M_S \rangle | n m_n \bar{\lambda} \epsilon \rangle | i_d
 m_{i_{d}} \rangle | i_a m_{i_{a}} \rangle
| L M_L \rangle.
\end{eqnarray}

The third basis set, B3, is intermediate between B1 and B2. Here
the molecule is again labeled by quantum numbers corresponding to
a Hund's case (a) molecule, as in B1,
 \begin{eqnarray}
 | B3w \rangle = |s_d \sigma \rangle | \lambda \rangle|j m \omega \rangle | i_d
 m_{i_{d}} \rangle |s_a m_{s_a} \rangle | i_a m_{i_a} \rangle| L M_L\rangle.
\end{eqnarray}
However, the part describing the OH fragment is taken as $|j m
\omega \rangle  | \lambda \rangle |s_d \sigma \rangle$, that is,
with signed values of $\lambda$, $\sigma$, and $\omega$. In this
case, the parity operator $E^{*}$ acts as follows,
\begin{align} &E^{*} |s_d \sigma \rangle | \lambda \rangle|j m \omega \rangle | i_d
 m_{i_{d}} \rangle |s_a m_{s_a} \rangle | i_a m_{i_{a}} \rangle | L M_L\rangle
 \nonumber \\ &= (-1)^{j-s_d+L} \nonumber \\ &\times |s_d -\sigma \rangle | -\lambda \rangle|j m -\omega \rangle | i_d
 m_{i_{d}} \rangle |s_a m_{s_a} \rangle | i_a m_{i_{a}} \rangle| L
 M_L\rangle,
\end{align}
so that we can again build combinations of defined parity,
 \begin{eqnarray}
 | B3p \rangle = |s_d \bar{\lambda} \bar{\omega} \epsilon j m \rangle | i_d
 m_{i_{d}} \rangle |s_a m_{s_a} \rangle | i_a m_{i_{a}} \rangle| L M_L\rangle.
\end{eqnarray}

The change from B2w to B3w, leaving aside the partial wave and
nuclear spin states, is given by
\begin{widetext}
 \begin{eqnarray}
 &|s_d \sigma \rangle& | \lambda \rangle| j m \omega \rangle |s_a
 m_{s_a}
\rangle \nonumber \\ &=&  \sum_{m_{s_d}} \sum_{n} \sum _{S} \sum
_{M_S} \left(\begin{array}{ccc} s_d & s_a &
 S \\ m_{s_d} & m_{s_a} & -M_S\end{array}\right)
\left(\begin{array}{ccc} s_d & n & j \\ m_{s_d} & m_n & -m
\end{array} \right) \left(\begin{array}{ccc} s_d & n & j \\ -
\sigma & -\lambda & \omega
 \end{array}
\right) \\ &\times& (-1)^{(m- \omega + s_d - s_a +M_S)} | n m_n
\lambda \rangle | (s_d s_a) S M_S \rangle \nonumber.
\end{eqnarray}
\end{widetext}

\end{document}